\documentclass[aps,prb,groupedaddress,twocolumn]{revtex4-2}

\usepackage{graphicx}
\usepackage{amsfonts}
\newcommand{\mycomment}[1]{}

\begin{document}


\title{Analysis of an all-to-all connected star array of transmon qubits}


\author{Ricardo A. Pinto}
\affiliation{School of Arts and Sciences, Massachusetts College of Pharmacy and Health Sciences, Boston, Massachusetts 02115, USA}


\date{\today}

\begin{abstract}
We analyzed quantum $XX$ and $ZZ$ coupling and state transfer in an all-to-all connected star array of capacitively coupled superconducting transmon qubits. It is shown that in a highly-connected system like this a variety of different $ZZ$ couplings arise that correspond to the different ways qubits can interact with each other, opening different channels for unwanted qubit crosstalk and thus qubit operation errors. We studied the dependence of both the $XX$ and the $ZZ$ coupling on qubit detuning that controls qubit-qubit interaction. The $XX$ coupling, quantified by the error state occupation probability, shows a $\Delta\omega^{-2}$ decay with qubit detuning $\Delta\omega$. On the other hand, all $ZZ$ coupling frequencies show spikes at values in the lower detuning region that correspond to resonances between qubit states and states out of the computational basis, after which all couplings quickly decay to zero as qubit detuning further increases. This allows to define an operational region where near-zero qubit coupling can be achieved. We derive equations for the couplings as a function of qubit detuning that agree with numerical results solving the Schr\"odinger equation.
\end{abstract}


\maketitle

\section{Introduction}


Architectures with high qubit connetivity have been proposed to provide a path to near-term pre-threshold quantum computing: They can simplify quantum gates for computation speedup and can be used for quantum simulations \cite{Geller2015, Katabarwa2015}, have applications in quantum error correction with lower overhead \cite{Bravyi2024, Xu2024}, and can potentially be used as a testbed for exploring quantum many-body physics in highly connected systems \cite{Xu2024}. Because of their high tunability and potential scalability, superconducting qubits have been used to explore the use of high-connectivity architectures for quantum computation, where simple graph circuit designs of three or four qubits where coupled in a star array to implement multi-qubits gates and demonstrate multi-qubits entanglement \cite{Neeley2010, Roy2020, Wu2024}.

Here we analyze an all-to-all connected star array of transmon qubits like the array implemented by Neeley {\it et al} using phase qubits \cite{Neeley2010} and study the dependence of the quantum qubit coupling with the parameters of the system. The coupling between qubits in this system is of capacitive type, which is the simplest coupling implementation and can potentially be upgraded to a design with tunable coupling. Nevertheless, it allows us to explore the fate of qubit-qubit interactions in the presence of higher qubit connectivity.

Notice that in this system the third qubit is not used as a coupler and therefore not always detuned from the other two qubits during qubit operations. All three qubits can be resonant with each other and this adds more complexity into the system, in particular when considering the $ZZ$ coupling between qubits that leads to qubit crosstalk. We show that there is not only pairwise $ZZ$ coupling between qubits, which can become significant when there is resonant interaction between qubit states and higher-energy states outside of the computational basis, but there is also a three-qubits (all-to-all) $ZZ$ coupling linking the frequency of one qubit to the state of the other two. This all-to-all $ZZ$ coupling can be even larger than its pairwise counterparts.

In Section \ref{systemhamiltonian} we describe the Hamiltonian of the system, defining the pairwise (two-qubits) $XX$ and $ZZ$ coupling frequencies, and introducing the all-to-all $ZZ$ coupling frequency. In Section \ref{quantumdegen} we quantum-mechanically analyze the system in the case where all qubits are degenerate (have the same qubit parameters and qubit frequencies) to lowest order perturbation theory using dressed states, and obtain the pairwise $XX$ and $ZZ$ coupling frequencies and the all-to-all $ZZ$ coupling frequency. In Section \ref{quantumdetuned} we do a similar analysis to characterize qubit $XX$ and $ZZ$ couplings when all qubits are detuned, where we use the error state occupation probability rather than the frequency to quantify the $XX$ coupling. We show that the error state occupation probability has a power-law decay when qubit detuning is large compared to the characteristic qubit coupling strength; and that the $ZZ$ couplings, before ultimately decaying to zero with qubit detuning, show spikes due to resonant interaction of qubit states with states outside of the computational basis. This helps to define a minimum detuning for attaining very low (OFF) coupling that allows for operations on individual qubits. Conclusions are in Section \ref{conclusions}, and in the Appendix we provide additional details of the analysis and explicit mathematical derivations.

\section{System and Hamiltonian}\label{systemhamiltonian}

\begin{figure}
\includegraphics[scale=0.28]{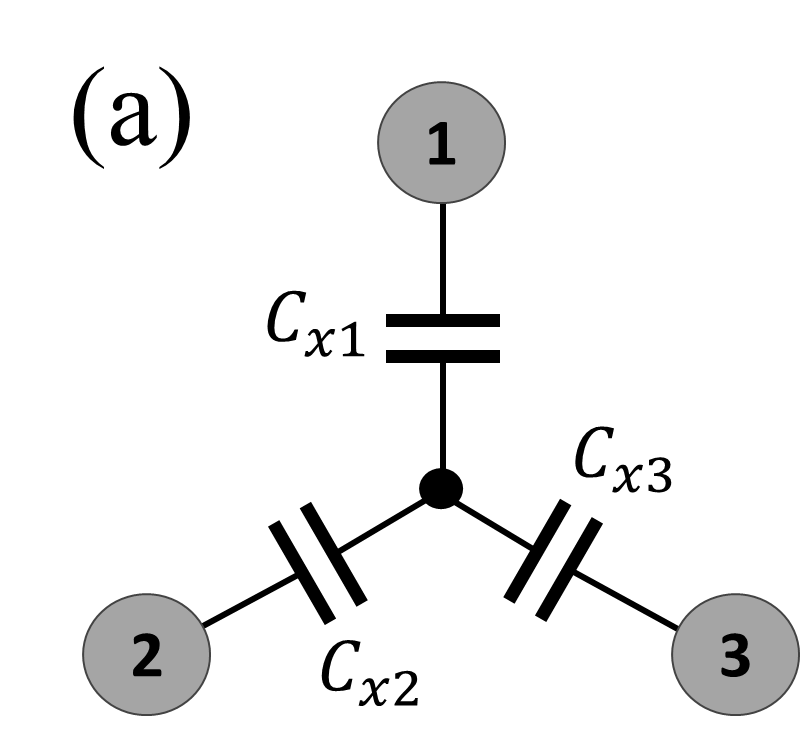}
\includegraphics[scale=0.28]{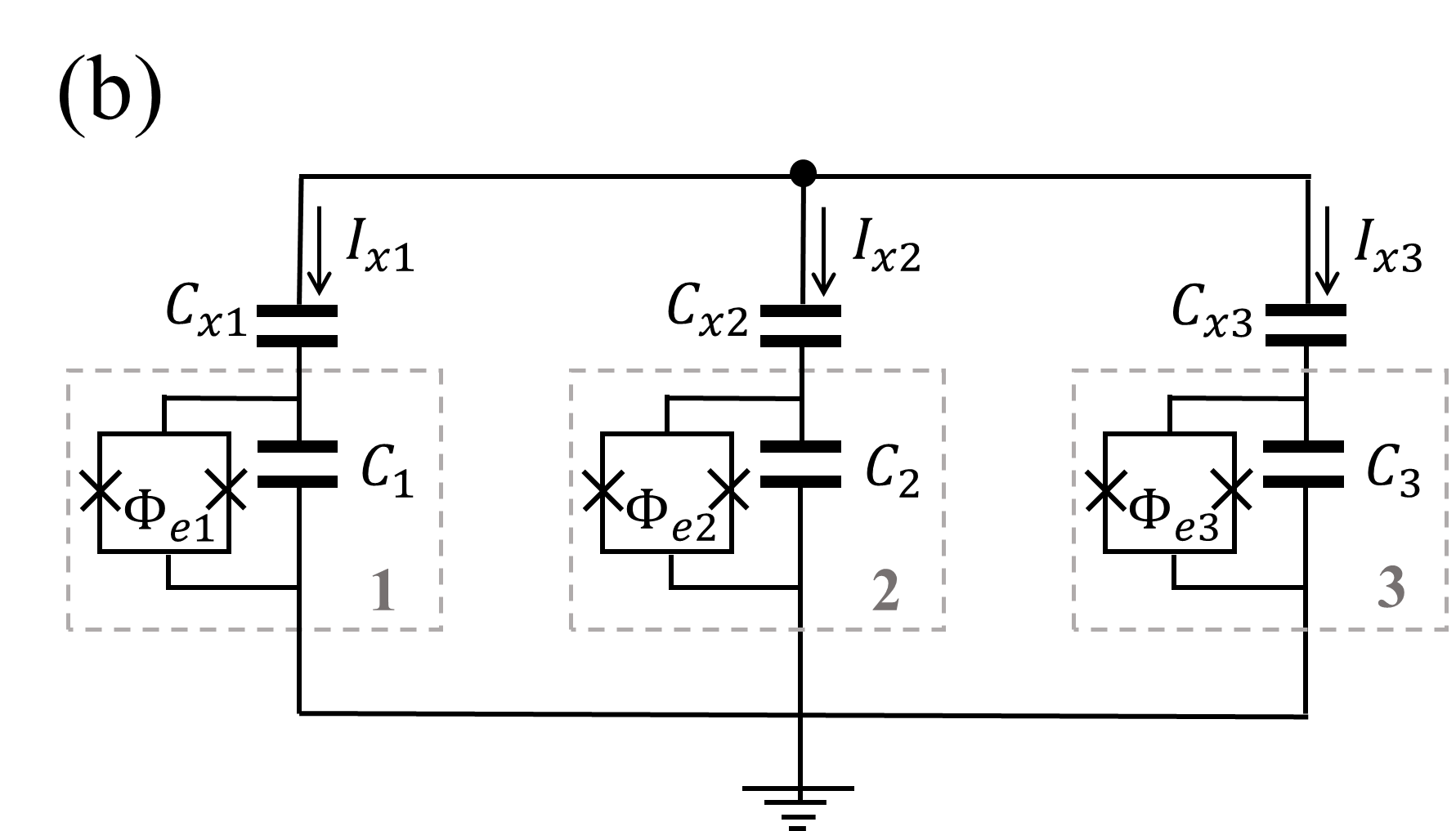}
\caption{(a) Circuit schematics of the system of three transmon qubits that are capacitively coupled in a star array, and (b) equivalent circuit, where the labeled gray dashed boxes contain the qubits represented as gray circles in (a). The current through each coupling capacitor $C_{xi}$ ($i=1,2,3$) is $I_{xi}$, and each qubit is biased by corresponding external magnetic flux $\Phi_{ei}$.\label{system}}
\end{figure}

The circuit design is displayed in Fig.\ref{system}-a. Three flux-biased transmon qubits \cite{Koch2007} are capacitively coupled to a central common island in a star array via the capacitors $C_{xi}\, (i=1,2,3)$. This circuit design was implemented with phase qubits by Neeley {\it et al} \cite{Neeley2010}. For easier analysis, this circuit can be redrawn into an equivalent circuit shown in Fig.\ref{system}-b. The Hamiltonian can be written as (see Appendix \ref{hamiltonian} for derivations) 
\begin{equation}
	H = \sum_{i=1}^{3} H_i + H_{int}, \label{eq:H}
\end{equation}
where
\begin{eqnarray}
	H_i &=& \frac{1}{2} \left(\frac{\Phi_0}{2\pi}\right)^2 \left[\mathbf{C^{-1}}\right]_{ii} p_i^2 + U_i(\delta\varphi_i), \\
	H_{int} &=&  \sum_{i=1}^{3}\sum_{j>i}^{3}  \left(\frac{\Phi_0}{2\pi}\right)^2 \left[\mathbf{C^{-1}}\right]_{ij} p_ip_j.
\end{eqnarray}
such the potentials $U_i$ have minima at $\delta\varphi_i = 0$. Notice that the system is invariant under cyclic qubit label permutations ($1\rightarrow 2\rightarrow  3\rightarrow 1$). $\mathbf{C}$ is the $3\times3$ capacitance matrix of the system, with matrix elements
\begin{eqnarray}
	C_{ii} &=& C_i + \frac{C_{xi}}{C_{\Sigma}}\tilde{C}_{xi},\\
	C_{ij} &=& \frac{C_{xi}C_{xj}}{C_{\Sigma}}, \, j\neq i,
\end{eqnarray}
where $C_{\Sigma} = \sum_{i=1}^{3}C_{xi}$ and $\tilde{C}_{xi} = \sum_{j\neq i}C_{xj}$.

Let $\left\{|n_i\rangle\right\}$ be the eigenstates of each individual qubit Hamiltonian $H_i$ in the absence of interaction with corresponding eigenvalues $\left\{\epsilon_{n_i}\right\}$, where the two lowest states $|0\rangle$ and $|1\rangle$ are used as logic states, and $|n_1n_2n_3\rangle = |n_1\rangle\otimes|n_2\rangle\otimes|n_3\rangle$ an eigenstate of the non-interacting three-qubits system with eigenvalue $\epsilon_{n_1n_2n_3} = \epsilon_{n_1} + \epsilon_{n_2} + \epsilon_{n_3}$. 

Similar to \cite{Pinto2010}, we define the coupling frequencies in terms of the exact eigenstates of the full physical system where the three-qubits logic states $|000\rangle$ and $|111\rangle$, for instance, are associated to the corresponding eigenstates of the full interacting Hamiltonian (\ref{eq:H}) with energies $E_{000}$ and $E_{111}$. 

In the absence of interaction, $\omega_{110} = \omega_{100} + \omega_{010}$, where $\hbar\omega_{n_1n_2n_3} = \epsilon_{n_1n_2n_3} - \epsilon_{000}$. In the interacting system, the eigenvalues $E_{n_1n_2n_3}$ are such that $\omega_{110}^{\prime} = \omega_{100}^{\prime} + \omega_{010}^{\prime} + \zeta_{110} \neq \omega_{100}^{\prime} + \omega_{010}^{\prime}$, where $\hbar\omega_{n_1n_2n_3}^{\prime} = E_{n_1n_2n_3} - E_{000}$. The difference $\zeta_{110}$ is the (pairwise) $ZZ$-coupling between qubits 1 and 2, and can be calculated from the eigenstates as \cite{Pinto2010, Peterson2024, DiCarlo2009}:
\begin{equation}
	\hbar\zeta_{110} = E_{110} + E_{000} - E_{100} - E_{010}, \label{eq:zz110}
\end{equation}
The coupling $\zeta_{110}$ is dominated by the interaction between the state $|110\rangle$ and nearly resonant states $|200\rangle$ and $|020\rangle$ that are out of the computational basis. 

Similarly, there is a $ZZ$ coupling $\zeta_{101}$ from interaction between the state $|101\rangle$ and states $|200\rangle$ and $|002\rangle$, and a $ZZ$ coupling  $\zeta_{011}$ from the interaction between the state $|011\rangle$ and states $|020\rangle$ and $|002\rangle$:
\begin{eqnarray}
	\hbar\zeta_{101} &=& E_{101} + E_{000} - E_{100} - E_{001}, \label{eq:zz101}\\
	\hbar\zeta_{011} &=& E_{011} + E_{000} - E_{010} - E_{001}, \label{eq:zz011}
\end{eqnarray}
which reflect the cyclic permutation symmetry of the system.

A third all-to-all $ZZ$ coupling linking the three qubits exists such that $\omega_{111}^{\prime} = \omega_{100}^{\prime} + \omega_{010}^{\prime} + \omega_{001} + \zeta_{111}$:
\begin{equation}
	\hbar\zeta_{111} = E_{111} + 2E_{000} - E_{100} - E_{010} - E_{001}, \label{eq:zz111}
\end{equation}
which would be dominated by the interaction between the state $|111\rangle$ and the higher-energy states $|300\rangle,\, |030\rangle,\, |003\rangle,\, |210\rangle,\, |021\rangle,\, |102\rangle,\, |201\rangle,\, |120\rangle$, and $|012\rangle$.

For a 4-qubits system there would be $4 \choose 2$ pairwise, $4 \choose 3$ three-qubits, and one four-qubits (all-to-all) $ZZ$ couplings; and for a $n$-qubits system there would be $n \choose 2$ pairwise, $n \choose 3$ three-qubits, $\ldots$, $n \choose n-1$ $n-1$-qubits, and one $n$-qubits (all-to-all) $ZZ$ couplings, giving a total of
\begin{equation}
	N_{zz} = \sum_{k=1}^{n-1} {n \choose k-1} + 1
\end{equation}
$ZZ$ couplings.
 
The quantum $XX$ coupling is defined as the minimum energy splitting in the avoided level crossing between the states $|100\rangle$, $|010\rangle$, and $|001\rangle$:
\begin{equation}
	\Omega_{XX} = \min_{\Delta\omega \to 0} \frac{1}{\hbar}|E_{+} - E_-|, \label{eq:XXcoupling},
\end{equation}
where $\Delta\omega$ is the detuning between qubits.

\section{Quantum analysis --degenerate qubits}\label{quantumdegen}

For the quantum analysis we use perturbation theory to derive the dressed states starting from the eigenstates $|n_1n_2n_3\rangle$ of the non-interacting Hamiltonian $\sum_{i=1}^{3}H_i$ \cite{Pinto2010}. The interaction Hamiltonian can be written in terms of the bosonic creation and annihilation operators $a,a^{\dagger}$ as
\begin{equation}
	H_{int} = -\sum_{i=1}^{3}\sum_{j>i}^{3} K_{ij}(a_i - a_i^{\dagger})(a_j - a_j^{\dagger}),
\end{equation}
where $a_i + a_i^\dagger = \delta\varphi_i\sqrt{2m_{i}\omega_i/\hbar}$ and $a_i - a_i^\dagger = \imath p_i\sqrt{2/(\hbar m_{i}\omega_i)}$. The interaction amplitudes are
\begin{equation}
	K_{ij} = \frac{\hbar \sqrt{m_im_j\omega_i\omega_j}}{2M_{ij}},\,\, j \neq i,
\end{equation}
where
\begin{eqnarray}
	M_{ij} &=&  \left(\frac{\Phi_0}{2\pi}\right)^2 \frac{1}{\left[\mathbf{C^{-1}}\right]_{ij}}, \\
	m_i &=& M_{ii}.
\end{eqnarray}
The qubit frequency $\omega_i$ is given by the energy difference between the two lowest qubit states, which due to anharmonicity is slightly smaller than the qubit plasma frequency $\omega_{pl,i}$ that depends on the qubit critical current $I_{c,i}$ (see Appendix \ref{hamiltonian}).

In the dressed states approach \cite{Pinto2010}, one expresses the solution of the Schr\"odinger equation $H|\psi\rangle = E|\psi\rangle$ as
\begin{equation}
	|\psi\rangle = \alpha |\psi_{100}^{dr}\rangle + \beta |\psi_{010}^{dr}\rangle + \gamma |\psi_{001}^{dr}\rangle, \label{eq:psi}
\end{equation}
The dressed state $|\psi_{100}^{dr}\rangle$ expanded in the product-state basis $\left\{|n_1n_2n_3\rangle\right\}$ has the contribution from the state $|100\rangle$ with amplitude 1 and zero contribution from the states $|010\rangle$ and $|001\rangle$, i.e., $\langle\psi_{100}^{dr}|100\rangle = 1$, $\langle\psi_{100}^{dr}|010\rangle = \langle\psi_{100}^{dr}|001\rangle = 0$. $|\psi_{100}^{dr}\rangle$ also satisfies the equation $\langle n_1n_2n_3|H|\psi_{100}^{dr}\rangle = E\langle n_1n_2n_3|\psi_{100}^{dr}\rangle$ for all basis elements $|n_1n_2n_3\rangle$ except $|100\rangle$, $|010\rangle$, and $|001\rangle$. The dressed states $|\psi_{010}^{dr}\rangle$ and $|\psi_{001}^{dr}\rangle$ are similarly defined, except that $\langle\psi_{010}^{dr}|010\rangle = 1$, $\langle\psi_{010}^{dr}|100\rangle = \langle\psi_{010}^{dr}|001\rangle = 0$ for $|\psi_{010}^{dr}\rangle$, and $\langle\psi_{001}^{dr}|001\rangle = 1$, $\langle\psi_{010}^{dr}|001\rangle = \langle\psi_{001}^{dr}|100\rangle = 0$ for $|\psi_{001}^{dr}\rangle$. Notice that a dressed state is not a solution of an eigenvalue problem; for a given energy $E$ it is a solution to an inhomogeneous system of linear equations. Also notice that we do not need to normalize the wavefunctions.

With the dressed states constructed above, we have only three equations to be self-consistently satisfied in order to solve the Schr\"odinger equation:
\begin{eqnarray}
	E_{100}^{dr}\alpha + V_{12}^{dr}\beta + V_{13}^{dr}\gamma &=& E\alpha, \label{eq:alpha} \\
	V_{21}^{dr}\alpha + E_{010}^{dr}\beta + V_{23}\gamma &=& E\beta,  \label{eq:beta} \\	
	V_{31}\alpha + V_{32}\beta + E_{001}^{dr}\gamma &=& E\gamma  \label{eq:gamma},
\end{eqnarray} 
where $E_{100}^{dr} \equiv \langle 100|H|\psi_{100}^{dr}\rangle$, $E_{010}^{dr} \equiv \langle 010|H|\psi_{010}^{dr}\rangle$, and $E_{001}^{dr} \equiv \langle 001|H|\psi_{001}^{dr}\rangle$ are the renormalized three-qubits self-energies and the effective interactions are $V_{12}^{dr} \equiv \langle 100|H|\psi_{010}^{dr}\rangle$, $V_{21}^{dr} \equiv \langle 010|H|\psi_{100}^{dr}\rangle$, $V_{13}^{dr} \equiv \langle 100|H|\psi_{001}^{dr}\rangle$, $V_{31}^{dr} \equiv \langle 001|H|\psi_{100}^{dr}\rangle$, $V_{23}^{dr} \equiv \langle 010|H|\psi_{001}^{dr}\rangle$, and $V_{32}^{dr} \equiv \langle 001|H|\psi_{010}^{dr}\rangle$.

To lowest order in perturbation theory, and neglecting terms with four or more total qubits excitation number, the dressed state $|\psi_{100}^{dr}\rangle$ and self-energy $E_{100}^{dr}$ are:
\begin{eqnarray}
	|\psi_{100}^{dr}\rangle &=& |100\rangle - \frac{\sqrt{2}K_{12}}{E - \epsilon_{210}}|210\rangle - \frac{\sqrt{2}K_{13}}{E - \epsilon_{201}}|201\rangle, \label{eq:psi100} \\
	E_{100}^{dr} &=& \epsilon_{100} + \frac{2K_{12}^2}{E - \epsilon_{210}} + \frac{2K_{13}^2}{E - \epsilon_{201}} + \frac{K_{23}^2}{E - \epsilon_{111}}.  \label{eq:E100}
\end{eqnarray}
The effective interaction $V_{12}^{dr} = V_{21}^{dr}$ is (see Appendix \ref{dressed} for the other effective interactions)
\begin{equation}
	V_{12}^{dr} = K_{12} + \frac{K_{13}K_{23}}{E - \epsilon_{111}}. \label{eq:V21}
\end{equation}
The other dressed states, self-energies, and effective interactions can easily be obtained by cyclic label permutations (see Appendix \ref{dressed} for explicit formulas).

For weak coupling ($C_{xi} \ll C_i$, $i=1,\,2,\,3$) we can approximate, for instance, $E \approx \epsilon_{100}$ in Eqs.(\ref{eq:psi100}-\ref{eq:V21}); and for degenerate qubits ($\omega_i = \omega_{qb}$) this leads to $E - \epsilon_{210} \approx E - \epsilon_{201} \approx E - \epsilon_{111} \approx -2\hbar\omega_{qb}$. 


For a symmetric system $C_i = C$ and $C_{xi} = C_x$ and hence $K_{ij} = -\hbar\omega_{qb}C_x/(6C)$. For this case $E_{100}^{dr} = E_{010}^{dr} = E_{001}^{dr} \equiv E_{[100]}^{dr}$ and $V_{12}^{dr} = V_{13}^{dr} = V_{23}^{dr} \equiv V^{dr}$. The $XX$ coupling becomes (see Appendix \ref{dressed} for derivations):
\begin{eqnarray}
	\Omega_{XX} &=& \frac{3}{\hbar}|V^{dr}| \\
	 &\simeq& \left(1 + \frac{C_x}{12C}\right)\frac{C_x}{C} \frac{\omega_{qb}}{2} \\ 
	 &\approx& \frac{C_x}{C} \frac{\omega_{qb}}{2}, \label{eq:OmegaXX}
\end{eqnarray}
which is consistent with results for phase qubits \cite{Neeley2010}. For typical experimental values  $I_{c[1,2,3]}=40\,\text{nA}$, $C = 100$ fF, $C_x = 1$ fF \cite{Yan2018}, and $\omega_{qb}/(2\pi) = 6$ GHz one obtains $\Omega_{XX}/(2\pi) \approx  30\,\text{MHz}$. 

\begin{figure}
	\includegraphics[scale=0.205]{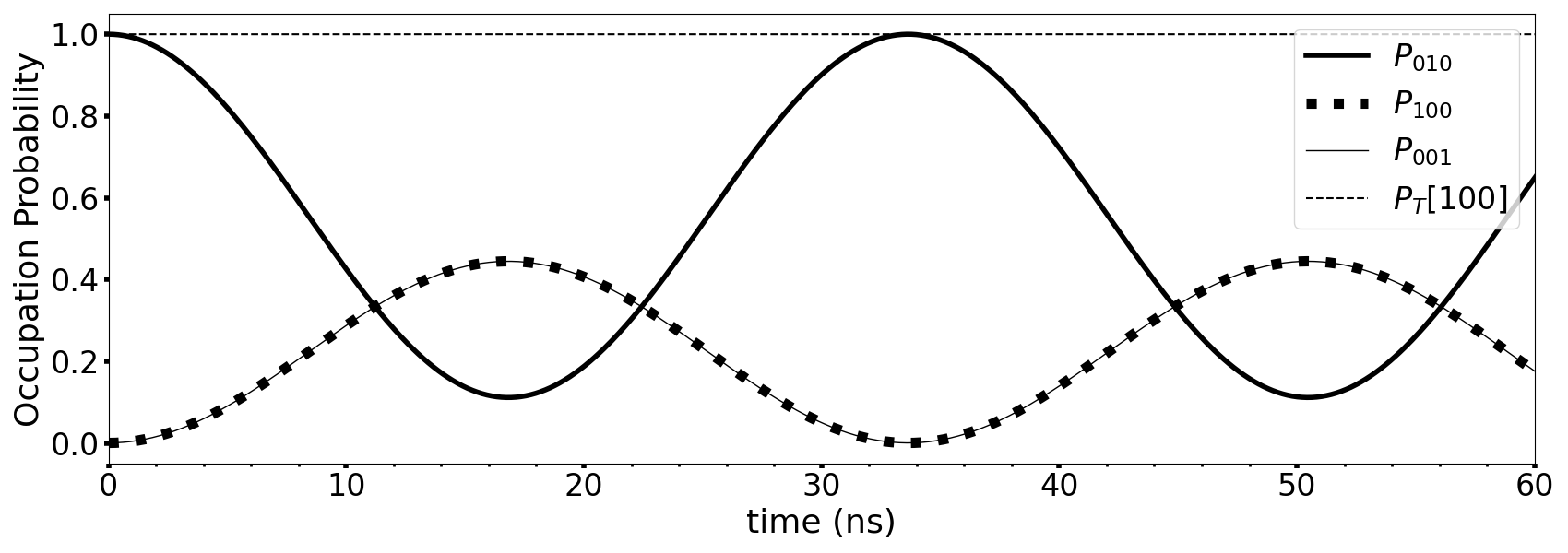}
	\caption{Time evolution of the single-excitation qubit state occupation probabilities for initial state $|\Psi_0\rangle = |010\rangle$. System's parameters are $I_{c[1,2,3]}= 40\,\text{nA},\,C_{1,2,3} = C= 100\,\text{fF},\, C_{x[1,2,3]} = C_x = 1\,\text{fF},\, \omega_{1,2,3} = \omega_{qb} = 6\,\text{GHz}$.\label{timedep010}}
\end{figure}

Fig. \ref{timedep010} shows the time evolution of the single-excitation qubit state occupation probabilities $P_{n_1n_2n_3}(t) = |\langle n_1n_2n_3|\Psi(t)\rangle|^2$ from numerical solution of the Schr\"odinger equation when the system starts with qubit 2 in the first excited state $|1\rangle$ and the other qubits are in their ground state $|0\rangle$. We see the expected equal distribution of the state occupation probability from qubit 2 to the other two qubits and then back to qubit 2 with an oscillation period of $\tau = 2\pi/\Omega_{XX} \simeq 33\,\text{ns}$ that corresponds to the 30 MHz $XX$ coupling we obtained above. The figure also shows the total single-excitation occupation probability $P_T[100] \equiv P_{100}+P_{010}+P_{001}$ that is nearly equal to one at all times, reflecting that the dynamics of the system remained, as expected, within the single-excitation subspace.
The evolution from the initial state $|010\rangle$ was previously used to generate the entangled state $|W\rangle = (|100\rangle + |010\rangle + |001\rangle)/\sqrt{3}$ \cite{Neeley2010}.

To obtain the four $ZZ$ couplings via Eqs. (\ref{eq:zz110}-\ref{eq:zz111}) we calculate $E_{000}$ as $E_{000}^{dr} = \langle 000|H|\psi_{000}^{dr}\rangle$, where we construct the dressed state $|\psi_{000}^{dr}\rangle$ as satisfying the Schr\"odinger equation $\langle n_1n_2n_3|H|\psi_{000}^{dr}\rangle = E\langle n_1n_2n_3|\psi_{000}^{dr}\rangle$ for all basis elements $|n_1n_2n_3\rangle$ except $|000\rangle$ and also satisfying $\langle 000|\psi_{000}^{dr}\rangle = 1$ assuming $E \approx \epsilon_{000}$. 

For the degenerate case there is ambiguity in assigning the eigenenergies $E_{110},\, E_{101}$, and $E_{011}$ to the corresponding bare states $|110\rangle,\,|101\rangle$, and $|011\rangle$; the same happens with $E_{100},\, E_{010}$, and $E_{001}$, so we cannot use each of the equations (\ref{eq:zz110}-\ref{eq:zz011}) separately to calculate the pairwise $ZZ$ couplings. The three-qubits $ZZ$ coupling, Eq. (\ref{eq:zz111}), can be directly used because $E_{111}$ unambiguously corresponds to the bare state $|111\rangle$ and the sum of the three first excited states $E_{100}+E_{010}+E_{001}$ does not need each separate eigenenergy to be unambiguously assigned to a specific bare state in the $\left\{|100\rangle,|010\rangle,|001\rangle \right\}$ subset.

To determine the pairwise coupling in the degenerate case we can, however, add Eqs. (\ref{eq:zz110}-\ref{eq:zz011}) and use the fact that the pairwise couplings should be equal to each other in the degenerate case, $\zeta_{110}=\zeta_{101}=\zeta_{011} = \zeta_p$. Thus
\begin{eqnarray}
\zeta_{p} = \frac{1}{3\hbar} \big[ &&E_{110}+E_{101}+E_{011} + 3E_{000} \nonumber \\
&& - 2\left(E_{100} + E_{010} + E_{001}\right) \big], \label{eq:ZZp}
\end{eqnarray}
where, similar to the first-to-third excited states discussed above, the sum of the seventh-to-ninth excited states $E_{110}+E_{101}+E_{011}$ does not need each separate eigenenergy to be unambiguously assigned to a bare state in the $\left\{|110\rangle,|101\rangle,|011\rangle \right\}$ subset.

Turning now to the calculation of individual eigenenergies in the degenerate-qubits case, the calculation of $E_{110}$ must account for the energy level interaction of the state $|110\rangle$ with the degenerate states $|101\rangle,|011\rangle$ and with the nearly degenerate states $|200\rangle, |020\rangle, |002\rangle$. Instead of writing a set of equations like (\ref{eq:alpha}-\ref{eq:gamma}) with all of these states, which would lead to six equations and thus to a sixth-degree equation for the eigenvalues,  we opt to separately add the contributions from these level interactions to $E_{110}^{dr} = \langle 000|H|\psi_{110}^{dr}\rangle$. The dressed state $|\psi_{110}^{dr}\rangle$ is constructed as satisfying the Schr\"odinger equation $\langle n_1n_2n_3|H|\psi_{110}^{dr}\rangle = E\langle n_1n_2n_3|\psi_{110}^{dr}\rangle$ for all basis elements $|n_1n_2n_3\rangle$ except $|110\rangle$, $|101\rangle$, $|011\rangle$, $|200\rangle$, $|020\rangle$, and $|002\rangle$; assuming $E \approx \epsilon_{110}$.

To lowest order in perturbation, the pairwise $ZZ$ coupling for degenerate qubits is (see Appendix \ref{couplingdegeneratederivation} for derivations)
\begin{eqnarray}
	\zeta_p \simeq  \omega_{qb} \Bigg[ && \sqrt{\alpha_{qb}^2 + 2\left(\frac{C_x}{3C}\right)^2} + \alpha_{qb} \nonumber \\ && + \frac{1}{2}\left(\frac{C_x}{3C}\right)^2 \left(2-\frac{1}{1+\alpha_{qb}}\right) \Bigg], \label{eq:zzp}
\end{eqnarray}
where $\alpha_{qb} = (\omega_{12} - \omega_{qb})/\omega_{qb} < 0$ is the qubit relative anharmonicity and $\omega_{12}$ is the $|1\rangle \rightarrow |2\rangle$ qubit transition frequency. In deriving Eq. (\ref{eq:zzp}) we accounted for the avoided level crossing between state $|110\rangle$ and states $|200\rangle$ and $|020\rangle$, which led to the first two terms. Also, the energy shifts in $E_{110},\,E_{101}$ and $E_{011}$ due to their avoided level crossing interaction cancel out in the sum $E_{110}+E_{101}+E_{011}$; the same happens with the interaction-induced energy shifts in $E_{100},\,E_{010}$ and $E_{001}$. The all-to-all three-qubits $ZZ$ coupling for degenerate qubits then is (see Appendix \ref{couplingdegeneratederivation})
\begin{equation}
	\zeta_{111}^{qb} \simeq \frac{3}{4} \left(\frac{C_x}{3C}\right)^2\omega_{qb} \left(\frac{4}{\alpha_{qb}} - \frac{5}{2}\right) \label{eq:ZZ111qb}.
\end{equation}

For the experimental parameters given earlier in the degenerate case, the pairwise $ZZ$ coupling $\zeta_{p}$ is around $1.89\,\text{MHz}$ from numeric calculations, and from Eq. (\ref{eq:zzp}) above it is $1.96\,\text{MHz}$. The $5\%$ difference can be accounted for by nonlinearity coefficients coming from the anharmonicity of the qubit potential that were neglected in the dressed state expansions. For this particular system's parameters at 6 GHz qubits frequency the qubit relative anharmonicity is $\alpha_{qb} \approx -0.035$, which in magnitude is only near ten times larger than the ratio $C_x/(3C)$. Thus the pairwise $ZZ$ coupling is only like fifteen times smaller than the $XX$ coupling, which is not negligible.

The all-to-all $ZZ$ coupling $\zeta_{111}^{qb}$ from analytics above is $5.6\,\text{MHz}$, which has a $16\%$ difference from numerical result of $4.7\,\text{MHz}$ that again is mostly accounted for nonlinearity coefficients that were neglected in the dressed state expansions. For the system's parameters above this coupling is only six times smaller than the $XX$ coupling, thus significant. Table \ref{tabledegenerate} summarizes the results for the degenerate case.
\begin{table} 
	\centering
	\caption{$XX$ and $ZZ$ coupling frequencies for degenerate qubits for typical experimental parameters  $I_{c[1,2,3]}=40\,\text{nA}$, $C_{1,2,3} = C = 100$ fF, $C_{x[1,2,3]} = C_x = 1$ fF, and $\omega_{1,2,3} = \omega_{qb}/(2\pi) = 6$ GHz. Both numerical (num) and analytical (ana) results are shown.}
	\label{tabledegenerate}
	\begin{tabular}{| c | c | c|}
		\hline
		\shortstack{$\Omega_{XX}/(2\pi)$ \\ (MHz)} & \shortstack{$\zeta_p/(2\pi)$ \\ (MHz)} & \shortstack{$\zeta_{111}^{qb}/(2\pi)$ \\ (MHz)} \\
		\hline
		\shortstack{30 (num) \\ 30 (ana)} & \shortstack{1.89 (num)\\ 1.96 (ana)} & \shortstack{4.7 (num) \\ 5.6 (ana)} \\
		\hline
	\end{tabular}
\end{table}

\section{Quantum analysis --detuned qubits}\label{quantumdetuned}

Knowing the effect of qubit detuning on the system's energy spectrum, and hence on the quantum coupling, is important when the qubits are brought out of resonance to effectively decouple them and perform operations on individual qubits.

To study the effect of detuning, for this three-qubits system we assume that qubit 1 and qubit 3 are detuned by the same amount $\Delta\omega \ll \omega_2$ below and above qubit-2 frequency $\omega_2$ respectively, i.e., $\omega_{1,3} = \omega_2\mp\Delta\omega$ \cite{Neeley2010}. Because the three qubits are now detuned, we can use Eqs. (\ref{eq:zz110}-\ref{eq:zz111}) to calculate the $ZZ$ couplings since we can unambiguously assign eigenenergies to corresponding bare states. 

In Eqs. (\ref{eq:alpha}-\ref{eq:gamma}), for weak qubit coupling $V_{ij}^{dr} \ll \hbar\omega_i$, $E_{100}^{dr} \simeq \epsilon_{100} = \hbar\omega_1 = \hbar\omega_2 - \Delta\omega$, $E_{010}^{dr} \simeq \epsilon_{010} = \hbar\omega_2$, and $E_{001}^{dr} \simeq \epsilon_{001} = \hbar\omega_3 =  \hbar\omega_2 + \Delta\omega$; and in the symmetric case the effective interactions $V_{ij}^{dr}$ become
\begin{eqnarray}
	V_{1[2,3]}^{dr} &\approx& K_{1[2,3]} = K\sqrt{1\mp\frac{\Delta\omega}{\omega_2}}, \\
	V_{23}^{dr} &\approx& K_{23} = K\sqrt{1-\left(\frac{\Delta\omega}{\omega_2}\right)^2},
\end{eqnarray}
where
\begin{equation}
	K = -\frac{C_x}{3C}\frac{\hbar\omega_2}{2}. \label{eq:K}
\end{equation}

We assume large qubit detuning, $\Delta\omega \gg |K|/\hbar$, where the eigenenergies are close to the system's energies in the non-interacting case, thus $E_{000} \simeq \epsilon_{000}$. The eigenenergy $E_{110}$ also approaches $\epsilon_{110}$ but we have to add the corrections due to the avoided level crossing interaction between $|110\rangle$ and the nearly degenerate states $|200\rangle,\,|020\rangle$, and $|002\rangle$.

\subsection{Error state occupation probability}

The XX coupling shown earlier was derived for the degenerate case. Defining the XX coupling in this three-qubits system when having detuning is more challenging. Instead, to quantify the coupling between qubits that are detuned we use the error state occupation probability
\begin{equation}
\mathcal{P}_e(010) = \frac{|\langle 100|\psi_{010}\rangle|^2 + |\langle 001|\psi_{010}\rangle|^2}{|\langle 100|\psi_{010}\rangle|^2 + |\langle 010|\psi_{010}\rangle|^2 + |\langle 001|\psi_{010}\rangle|^2 }
\end{equation}

The meaning of $\mathcal{P}_e(010)$ is the following: For large qubit detuning one expects that the eigenstate $|\psi_{010}\rangle$, which is close to the state $|010\rangle$, has a negligible contribution from the states $|100\rangle$ and $|001\rangle$ such that all three are decoupled. However, these contributions cannot be lowered to zero and $\mathcal{P}_e(010)$ measures the probability of either of the states $|100\rangle$ and $|001\rangle$ being wrongly populated when the system is in the eigenstate $|\psi_{010}\rangle$.

Solving Eqs. (\ref{eq:alpha}-\ref{eq:gamma}) for large detuning $\Delta\omega \gg K/\hbar$ but with $\Delta\omega \ll \omega_2$ leads us to the three eigenstates that are close to states $|100\rangle,\,|010\rangle$, and $|001\rangle$ in terms of their corresponding dressed states described after Eq. (\ref{eq:psi}) (see Appendix \ref{errorstatederivation} for derivations):
\begin{eqnarray}
	\psi_{100} &\simeq& |\psi_{100}^{dr}\rangle +  \frac{K}{\hbar\Delta\omega} \left(1-\frac{\Delta\omega}{2\omega_2}\right) |\psi_{010}^{dr}\rangle \nonumber \\
	& & + \frac{K}{2\hbar\Delta\omega} |\psi_{001}^{dr}\rangle, \label{eq:psi100detuned}\\
	\psi_{010} &\simeq& \frac{K}{\hbar\Delta\omega} \left(1+\frac{3}{2}\frac{\Delta\omega}{\omega_2}\right) |\psi_{100}^{dr}\rangle + |\psi_{010}^{dr}\rangle \nonumber \\
	& & -  \frac{K}{\hbar\Delta\omega} \left(1+\frac{\Delta\omega}{2\omega_2}\right) |\psi_{001}^{dr}\rangle, \label{eq:psi010detuned}\\
	\psi_{001} &\simeq& \frac{K}{\hbar\Delta\omega} |\psi_{010}^{dr}\rangle + \frac{K}{\hbar\Delta\omega}\left(1-\frac{\Delta\omega}{2\omega_2}\right) |\psi_{010}^{dr}\rangle \nonumber \\
	& & - |\psi_{001}^{dr}\rangle \label{eq:psi001detuned},
\end{eqnarray}
Thus, the error state occupation probability is:
\begin{equation}
	\mathcal{P}_e(010) \simeq \frac{1}{2}\left(\frac{C_x}{3C}\right)^2 \frac{\omega_2}{\Delta\omega}\left(\frac{\omega_2}{\Delta\omega}+2\right). \label{eq:pe}
\end{equation}
\begin{figure}
	\includegraphics[scale=0.31]{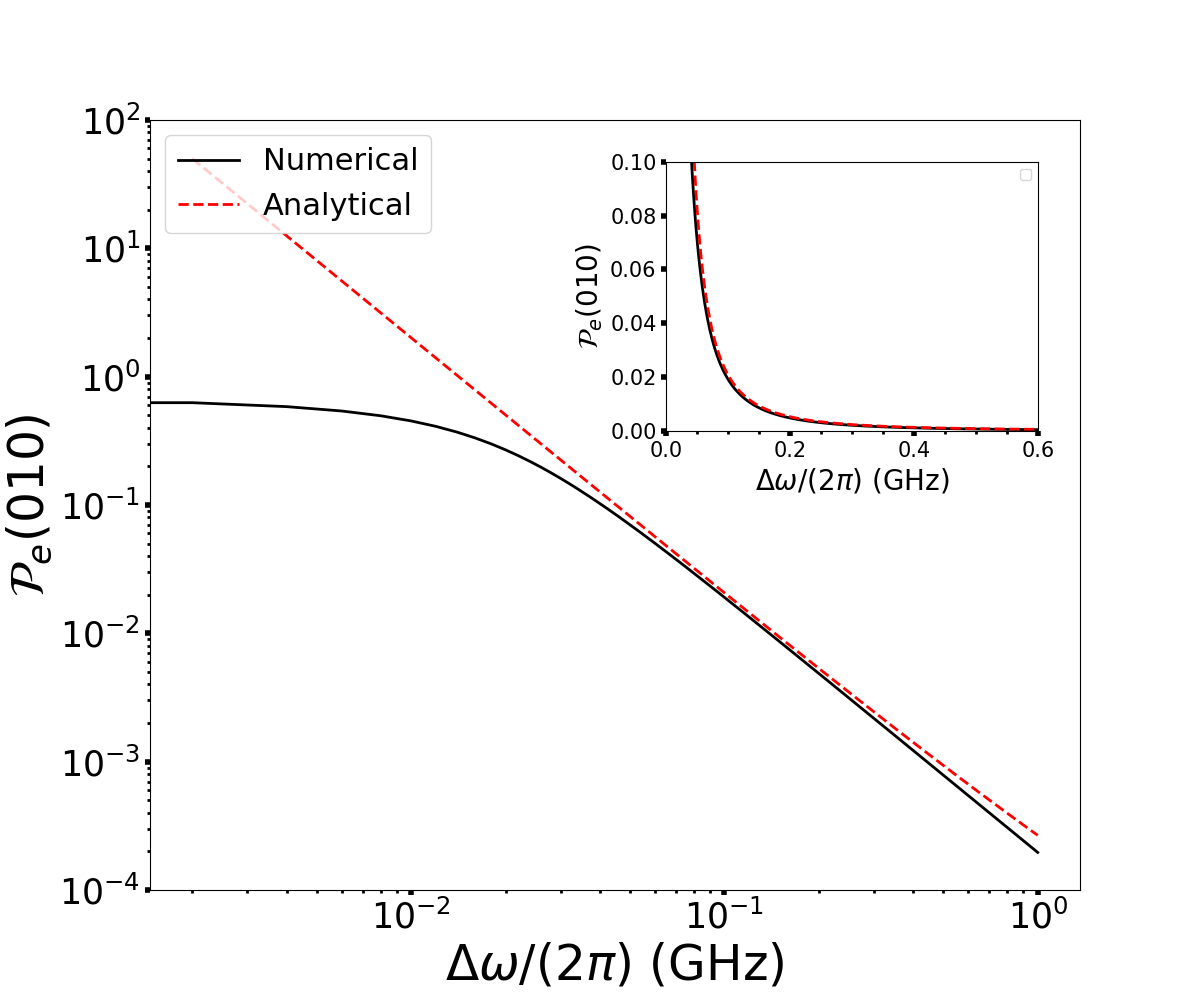}
	\caption{Error state occupation probability $\mathcal{P}_e(010)$ vs. qubit detuning $\Delta\omega$ in log-log scale. Black solid line: Numerical results. Red dashed line: Analytical results using Eq. (\ref{eq:pe}). Inset: Same plot in linear scale. Except for qubits 1 and 3 frequencies, which were detuned by $\mp\Delta\omega/(2\pi)$ from qubit 2 frequency of 6 GHz, all system's parameters are the same as in the degenerate case  (table \ref{tabledegenerate}).\label{pe010}}
\end{figure}

Fig. \ref{pe010} shows the error state occupation probability calculated from the numerically-obtained eigenstates of the system (black solid line). The system is symmetric with the same parameters as in the degenerate case, except that the magnetic fluxes through the qubits are such that $\omega_2/(2\pi) = 6\,\text{GHz}$ and $\omega_{1,3} = \omega_2\mp\Delta\omega$. The error state occupation probability shows a $\sim \Delta\omega^{-2}$ decay with qubits detuning described by Eq. (\ref{eq:pe}) for large detuning $\Delta$$\omega/(2\pi) \gg K/h \simeq 10\,\text{MHz}$  (red dashed line). The difference between analytical and numerical results starts to increase for $\Delta\omega/(2\pi) \simeq 1\,\text{GHz}$, where the approximation $\Delta\omega \ll \omega_2$ is no longer valid.

\subsection{ZZ coupling}

The correction to $E_{110}$ due to interaction with state $|200\rangle$ is given by the known formula describing an avoided crossing between to interacting states: $[\epsilon_{200}-\epsilon_{110} \pm \sqrt{(\epsilon_{200}-\epsilon_{110})^2 + S_{|110\rangle,|200\rangle}^2}]/2$, where $S_{|110\rangle,|200\rangle} = 2\langle 110|H|\psi_{200}^{dr}\rangle \approx 2\sqrt{2}K_{12}$ (see Appendix \ref{zzdetunedderivation} for derivations). The sign of the square-root term corresponds to whether the interaction with the state $|200\rangle$ pushes $E_{110}$ up ($\epsilon_{200} < \epsilon_{110}$) or down ($\epsilon_{200} > \epsilon_{110}$). Similarly, the corrections to $E_{110}$ due to interaction with states $|020\rangle$ and $|002\rangle$ are $[\epsilon_{020}-\epsilon_{110} \pm \sqrt{(\epsilon_{020}-\epsilon_{110})^2 + S_{|110\rangle,|020\rangle}^2}]/2$ and $[\epsilon_{002}-\epsilon_{110} \pm \sqrt{(\epsilon_{002}-\epsilon_{110})^2 + S_{|110\rangle,|002\rangle}^2}]/2$ respectively, where $S_{|110\rangle,|020\rangle} = 2\langle 110|H|\psi_{020}^{dr}\rangle \approx 2\sqrt{2}K_{12}$ and $S_{|110\rangle,|002\rangle}  = 2\langle 110|H|\psi_{002}^{dr}\rangle$ is a higher-order effective interaction mediated by the states $|101\rangle$ and $|011\rangle$: 
\begin{equation}
	S_{|110\rangle,|002\rangle} \approx 2\sqrt{2}K_{13}K_{23} \left(\frac{1}{\varepsilon_{1+} - \varepsilon_{1-}} + \frac{1}{\varepsilon_{1+}^{\prime} - \varepsilon_{1-}^{\prime}}\right), \label{eq:S110_002}
\end{equation}
The energies
\begin{equation}
\varepsilon_{1\pm} = \frac{1}{2}\left[ \epsilon_{002}+\epsilon_{101} \pm \sqrt{(\epsilon_{002}-\epsilon_{101})^2 + S_{|101\rangle,|002\rangle}^2} \right]
\end{equation}
replace the nearly degenerate (and interacting) energies $\epsilon_{002}$ and $\epsilon_{101}$, and the energies
\begin{equation}
\varepsilon_{1\pm}^{\prime} = \frac{1}{2}\left[ \epsilon_{002}+\epsilon_{011} \pm \sqrt{(\epsilon_{002}-\epsilon_{011})^2 + S_{|011\rangle,|002\rangle}^2} \right]
\end{equation}
replace $\epsilon_{002}$ and $\epsilon_{011}$; where $S_{|101\rangle,|002\rangle} = 2\langle 101|H|\psi_{002}^{dr}\rangle \approx 2\sqrt{2}K_{13}$ and $S_{|011\rangle,|002\rangle} = 2\langle 011|H|\psi_{002}^{dr}\rangle \approx 2\sqrt{2}K_{23}$.

Thus, the $ZZ$ coupling $\zeta_{110}$ is (see Appendix \ref{zzdetunedderivation} for derivations and the other $ZZ$ couplings):
\begin{eqnarray}
	\zeta_{110}(\Delta\omega) \simeq && \frac{\omega_2}{2} \left[\alpha_1 + \alpha_2 + \alpha_3 + (3-\alpha_1+\alpha_3)\frac{\Delta\omega}{\omega_2} \right]   \nonumber \\
	&& + \frac{\omega_2}{2} \mathcal{P}\left[\frac{(1+\alpha_1)\Delta\omega}{\omega_2} - \alpha_1; \frac{2K_{12}}{\hbar\omega_2}\right] \nonumber \\
	&& - \frac{\omega_2}{2} \mathcal{P}\left(\frac{\Delta\omega}{\omega_2}+\alpha_2; \frac{2K_{12}}{\hbar\omega_2}\right) \nonumber\\
	&& - \frac{\omega_2}{2} \mathcal{P}\left[\frac{(3+\alpha_3)\Delta\omega}{\omega_2} + \alpha_3; \frac{\mathcal{S}_{|110\rangle,|002\rangle}}{\sqrt{2}}\right], \label{eq:zeta110}
\end{eqnarray}
where
\begin{equation}
	\mathcal{P}(x; k) = \left[2\theta(x) - 1\right] \sqrt{ x^2 + 2k^2 }, \label{eq:P}
\end{equation}
$\theta(x)$ is the Heaviside function, and
\begin{widetext}
	\begin{equation}
		\mathcal{S}_{|110\rangle, |002\rangle} = \left(\frac{C_x}{3C}\right)^2 \frac{\sqrt{1+\frac{\Delta\omega}{\omega_2}}}{\sqrt{2}} \left\{ \frac{1}{\sqrt{\left[\frac{(2+\alpha_3)\Delta\omega}{\omega_2}+\alpha_3\right]^2 + 2\left(\frac{C_x}{3C}\right)^2}} 
		+ \frac{1}{\sqrt{\left[\frac{(1+\alpha_3)\Delta\omega}{\omega_2}+\alpha_3\right]^2 + 2\left(\frac{C_x}{3C}\right)^2\left(1+\frac{\Delta\omega}{\omega_2}\right) } } \right\}.
	\end{equation}
\end{widetext}

The first arguments in the three $\mathcal{P}$ functions in Eq. (\ref{eq:zeta110}) correspond to the energy differences $\epsilon_{200} - \epsilon_{110}$, $\epsilon_{020} - \epsilon_{110}$, and $\epsilon_{002} - \epsilon_{110}$ respectively, and $\alpha_{i} = [\omega_{12}^{(i)}-\omega_i]/\omega_i$ is the relative anharmonicity of qubit $i$. At the $\Delta\omega$ values for which they are equal to zero (each energy pair become resonant) the pairwise $ZZ$ coupling $\zeta_{110}$ in Eq. (\ref{eq:zeta110}) has corresponding spikes. 

\begin{figure}[b]
	\includegraphics[scale=0.34]{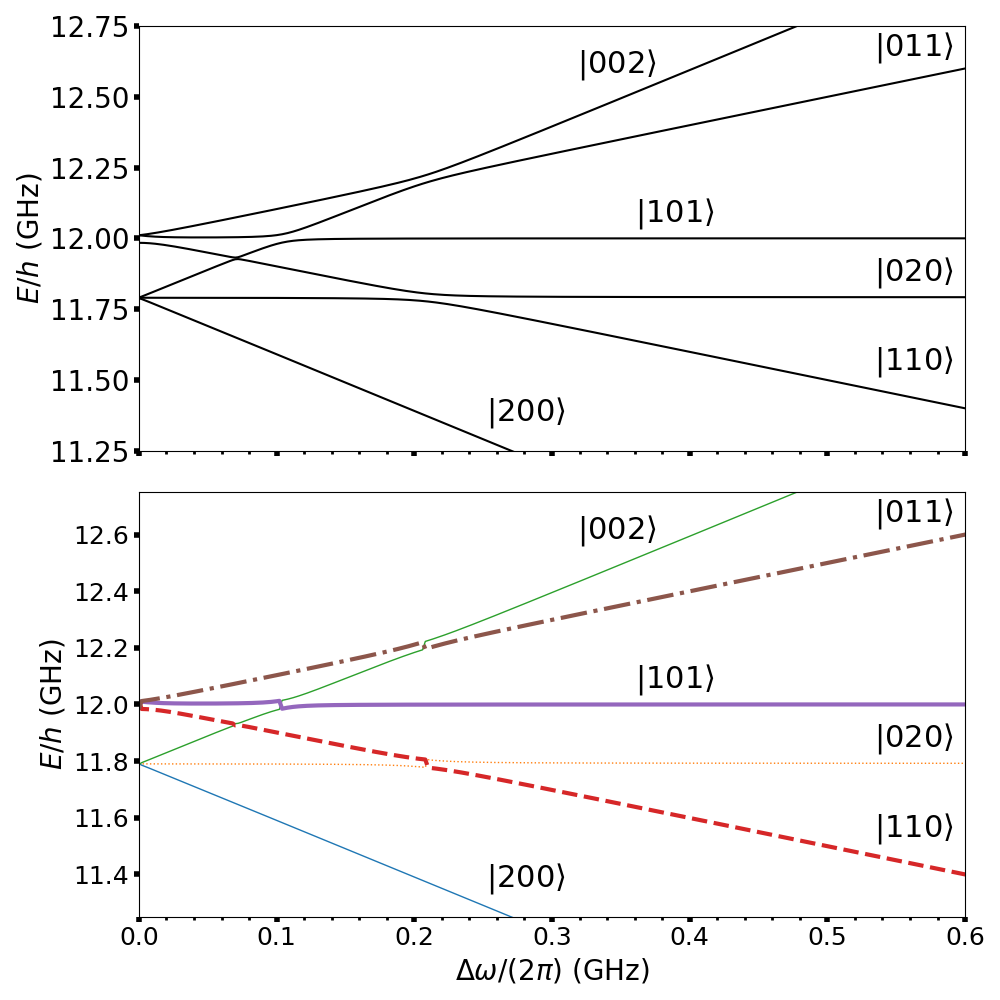}
	\caption{(Top) Energy spectrum of the system vs. qubit detuning in the two-excitation region, where the energy levels are labeled by the corresponding state. (Bottom) Same as the top panel but with the energy levels having different colors and line styles to track the corresponding state as detuning varies. Parameters are the same as in Fig. \ref{pe010}.\label{spectrum}}
\end{figure}

\begin{figure}
	\includegraphics[scale=0.37]{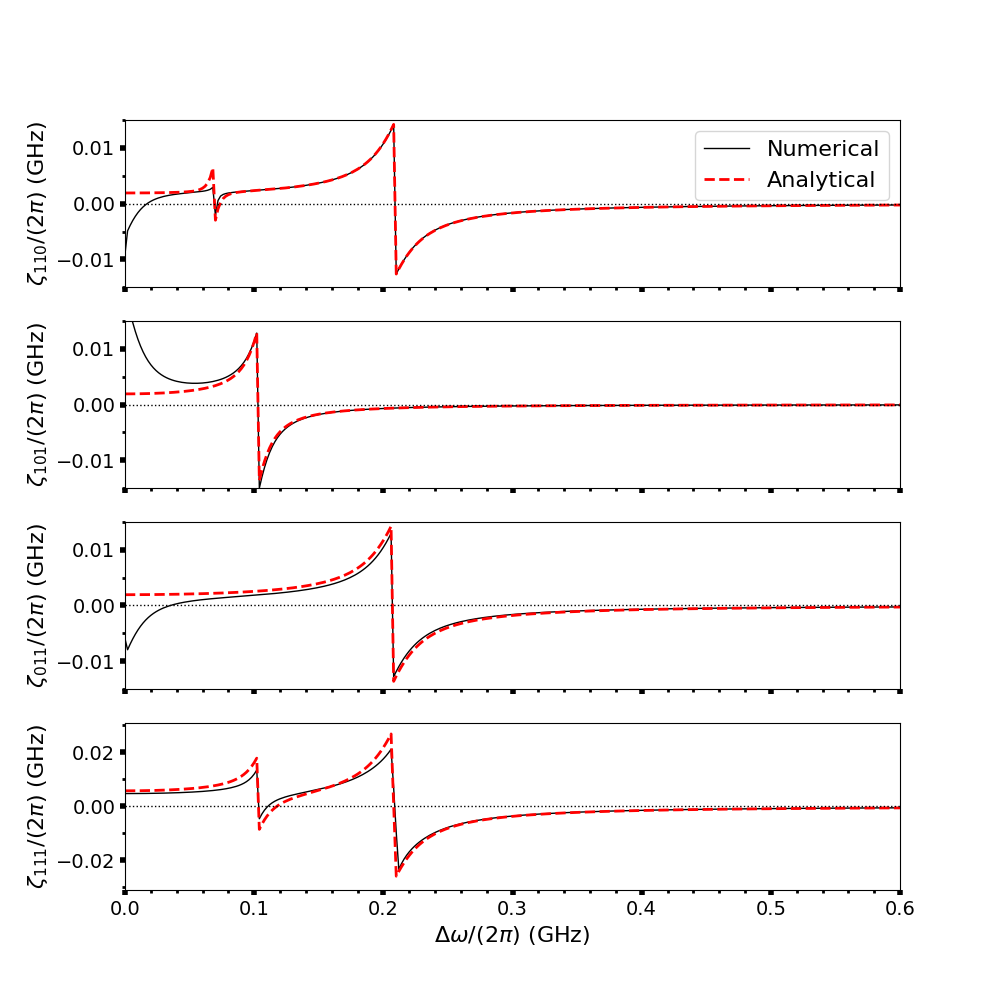}
	\caption{Pairwise $ZZ$ couplings vs. qubit detuning $\Delta\omega$.\label{zz}. Red dashed lines are analytical results using Eqs. (\ref{eq:zeta110}) and (\ref{eq:zeta101}-\ref{eq:zeta011}). The dotted line at zero coupling is a guide for the eye. Parameters are the same as in Fig. \ref{pe010}}
\end{figure}

The top panel of Fig. \ref{spectrum} shows the energy levels of the system in the two-excitation region as a function of qubit detuning. The bottom panel shows the same spectrum but this time we track the state corresponding to each energy level by using different line colors and styles. We see the avoided level crossings between the states $|110\rangle,\,|101\rangle,\,|011\rangle$ and the nearly degenerate states $|200\rangle,\,|020\rangle,\,|002\rangle$ that lead to the spikes in the pairwise $ZZ$ couplings as shown in the Fig \ref{zz}, which are well described by Eq. (\ref{eq:zeta110}).

The state $|110\rangle$ becomes resonant and has an avoided level crossing with state $|002\rangle$ at $\Delta\omega = -\alpha_3\omega_2/(3+\alpha_3)$, or 69 MHz for $\alpha_3 \simeq -0.034$ at that detuning value, and then with state $|020\rangle$ at $\Delta\omega = -\alpha_2\omega_2$ or 210 MHz ($\alpha_2 \simeq -0.035$ at that detuning value). Interaction with state $|200\rangle$ would require $\Delta\omega <0$ and thus the corresponding avoided crossing does not show up neither in Fig \ref{spectrum} nor in Fig. \ref{zz}. After the spikes, $\zeta_{110}$ approaches zero as detuning increases further.

Similarly, states $|101\rangle$ and $|011\rangle$ become resonant and have avoided level crossings with state $|002\rangle$ at $\Delta\omega = -\alpha_3\omega_2/(2+\alpha_3)$, or 104 MHz, and at $\Delta\omega = \alpha_3\omega_2/(1-\alpha_3)$, or 211 MHz, respectively and the pairwise $ZZ$ couplings $\zeta_{101}$ and $\zeta_{011}$ show corresponding spikes at those detuning values; after which both approach zero as detuning increases further. These behaviors are well described by Eqs (\ref{eq:zeta101} and \ref{eq:zeta011}).

Eq. (\ref{eq:zeta110}) for $\zeta_{110}$ (and the equations for the other pairwise $ZZ$ couplings) were obtained assuming large detuning and they nicely describe the behavior of the three $ZZ$ couplings when $\Delta\omega > |K|/\hbar = C_x\omega_2/(3C)$, or $\Delta\omega/(2\pi) > 20\,\text{MHz}$ as shown in Fig. \ref{zz}. For the chosen system's parameters the pairwise $ZZ$ couplings can reach a near 15 MHz peak at around $\Delta\omega/(2\pi) \simeq 200\, \text{MHz}$ detuning, about one order of magnitude larger than the near 2 MHz pairwise $ZZ$ coupling in the degenerate case at zero detuning we already explored.

Calculation of the all-to-all three-qubits $ZZ$ coupling $\zeta_{111}$ from Eq.(\ref{eq:zz111}) is similar to the pairwise case, where the eigenenergy $E_{111}$ approximates the bare energy $\epsilon_{111}$ and we have to add the corrections due to resonant interactions of the state $|111\rangle$ with states $|210\rangle,\,|021\rangle,\,|102\rangle,\,|201\rangle,\,|120\rangle,\,|012\rangle,\,|300\rangle,\,|030\rangle$, and $|003\rangle$ (see Appendix \ref{zzdetunedderivation} for derivations):
\begin{widetext}
	\begin{eqnarray}
		\zeta_{111} \simeq & & \omega_2 \left[ \alpha_1\left(1 - \frac{\Delta\omega}{\omega_2}\right) + \alpha_2 + \alpha_3\left(1+\frac{\Delta\omega}{\omega_2}\right)\right] +  \frac{\omega_2}{2} \mathcal{P}\left[\frac{(1+\alpha_1)\Delta\omega}{\omega_2} - \alpha_1; \frac{2K_{12}}{\hbar\omega_2}\right]
		+ \frac{\omega_2}{2} \mathcal{P}\left[\frac{(2+\alpha_1)\Delta\omega}{\omega_2} - \alpha_1; \frac{2K_{13}}{\hbar\omega_2} \right] \nonumber \\
		&& +  \frac{\omega_2}{2} \mathcal{P}\left(\frac{\Delta\omega}{\omega_2} - \alpha_2; \frac{2K_{23}}{\hbar\omega_2}\right)
	 	-  \frac{\omega_2}{2} \mathcal{P}\left(\frac{\Delta\omega}{\omega_2} + \alpha_2; \frac{2K_{12}}{\hbar\omega_2}\right)
		- \frac{\omega_2}{2} \mathcal{P}\left[\frac{(2+\alpha_3)\Delta\omega}{\omega_2} + \alpha_3; \frac{2K_{13}}{\hbar\omega_2} \right] \nonumber \\
		&& - \frac{\omega_2}{2} \mathcal{P}\left[\frac{(1+\alpha_3)\Delta\omega}{\omega_2} + \alpha_3; \frac{2K_{23}}{\hbar\omega_2} \right]. \label{eq:zeta111}
	\end{eqnarray}
\end{widetext}	

\begin{figure}
	\includegraphics[scale=0.34]{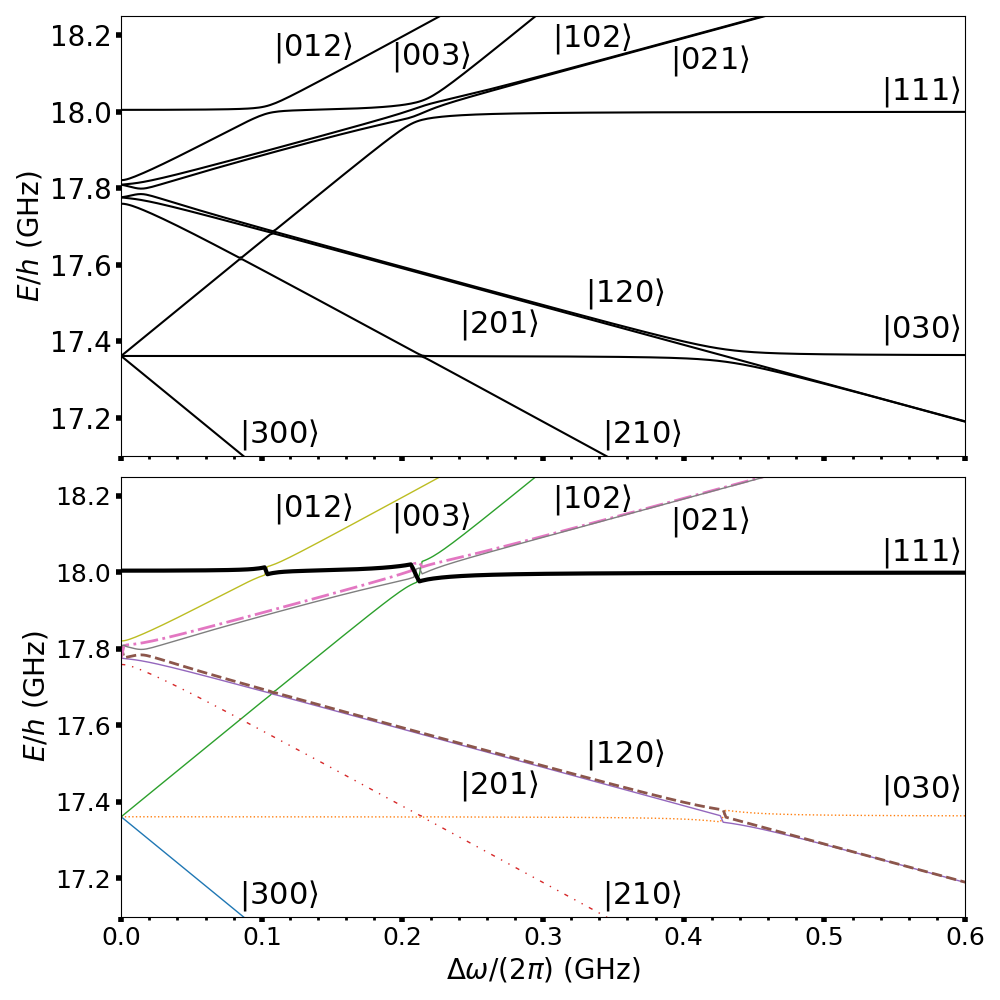}
	\caption{(Top) Energy spectrum of the system vs. qubit detuning in the three-excitation region, where the energy levels are labeled by the corresponding state. (Bottom) Same as the top panel but with the energy levels having different colors and line styles to track the corresponding state as detuning varies. Parameters are the same as in Fig. \ref{pe010}.\label{spectrum2}}
\end{figure}

After the first term in Eq. (\ref{eq:zeta111}), the first three $\mathcal{P}$-terms correspond to the interaction of state $|111\rangle$ with states $|201\rangle,\, |210\rangle$, and $|120\rangle$ respectively; which only occur when $\Delta\omega < 0$ and thus do not show up neither as avoided level crossings in the spectrum (Fig. \ref{spectrum2}) nor as spikes in the all-to-all $ZZ$ coupling $\zeta_{111}$ (Fig. \ref{zz}). The last three $\mathcal{P}$-terms correspond to the interaction of state $|111\rangle$ with states $|021\rangle,\, |012\rangle$, and $|102\rangle$ respectively; which occur for $\Delta\omega > 0$ and thus the corresponding avoided level crossings and $\zeta_{111}$ spikes show up in Figs. \ref{spectrum2} and \ref{zz} respectively. While the interaction with state $|012\rangle$ occurs at $\Delta\omega = -\alpha_3\omega_2/(2+\alpha_3)$, or 104 MHz ($\alpha_3 \simeq -0.0341$), the interaction with states $|021\rangle$ and $|102\rangle$ occurs at very close detuning values $\Delta\omega = \alpha_2\omega_2 \simeq -\alpha_3\omega_2/(1+\alpha_3)$ around 210 MHz ($\alpha_2 \simeq -0.0347$ and $\alpha_3 \simeq -0.0334$). The spikes in $\zeta_{111}$ at these detuning values are well described by Eq. (\ref{eq:zeta111}) as seen in Fig. \ref{zz}.

\begin{table} 
	\centering
	\caption{Location of the $ZZ$ coupling spikes vs qubit detuning. Parameters are the same as in Fig. \ref{pe010}.}
	\label{tableresonances}
	\begin{tabular}{| c | c | c | c |}
		\hline
		\multicolumn{4}{|c|}{Qubit detuning at the spikes $\Delta$$\omega/(2\pi)$ (MHz)} \\
		\hline
		$\zeta_{110}$ & $\zeta_{101}$ & $\zeta_{011}$ & $\zeta_{111}$ \\
		\hline
		69 & - & -	& - \\
		 - & 104 & - & 104 \\	
		210 & - & 211 & 210 \\
		\hline
	\end{tabular}
\end{table}

As visible in Fig. \ref{spectrum2}, the state $|111\rangle$ also interacts with the nearly degenerate state $|003\rangle$ at $\Delta\omega = -(2\alpha_3+\alpha_3^{\ast})/(3+2\alpha_3+\alpha_3^{\ast})$, or 212 MHz ($\alpha_3^{\ast} \simeq -0.0354$); where $\alpha^{\ast} = (\omega_{23}-\omega_{01})/\omega_{01}$ is the relative anharmonicity characterizing the transition frequency between qubit states $|2\rangle$ and $|3\rangle$. At that value of the qubits detuning $\epsilon_{003} = \epsilon_{111}$ and it is very close to where interaction of state $|111\rangle$ with states $|021\rangle$ and $|102\rangle$ occurs, combining together to produce a $\zeta_{111}$ that is larger compared to the pairwise $ZZ$ couplings (it can go over 20 MHz). The effective interaction with state $|003\rangle$ is of higher-order similar to Eq. (\ref{eq:S110_002}), mediated by states $|102\rangle$ and $|012\rangle$, and contributes little to the $\zeta_{111}$ and thus we did not include it in Eq. (\ref{eq:zeta111}). Higher-order interaction with state $|300\rangle$ (where $\epsilon_{300} = \epsilon_{111}$) occurs only if $\Delta\omega < 0$ and thus the corresponding avoided level crossing and $\zeta_{111}$ spike do not show up in Figs. \ref{spectrum2} and \ref{zz}; and interaction with state $|030\rangle$ never happens because the difference $\epsilon_{030} - \epsilon_{111}$ in independent of qubit detuning. Table \ref{tableresonances} summarizes the detuning values at which the spikes in $ZZ$ couplings occur.

\section{Conclusions}\label{conclusions}

We used dressed states and perturbation theory to analyze a system of three capacitively coupled transmon qubits connected in an all-to-all star array, and studied the dependence of the qubits effective coupling on the system's parameters, where the system is symmetric under cyclic qubit label permutations.

Different from most studies of coupling in three-qubits systems where one of them is used as a coupler and its frequency is different from the other two qubits, in this system all three qubits can be resonant to each other. For degenerate qubits, the single-excitation subspace in the energy spectrum has three degenerate levels that get partially lifted by the qubit interaction, where there is a single lower energy level and two still-degenerate higher levels. The difference between them defines the $XX$ coupling between the single-excitation states $|100\rangle,\,|010\rangle$, and $|001\rangle$, which for typical transmon parameters and small coupler capacitance can reach a few tens MHz.

We showed that having more than two coupled qubits introduces more complexity in the $ZZ$ coupling. For the specific case of a three-qubits system there are two types of $ZZ$ couplings: Three {\it pairwise} $ZZ$ coupling linking the frequency of one qubit to the state of one of the other two and one {\it all-to-all} three-qubits $ZZ$ coupling linking the frequency of one qubit to the state of all other qubits. For the degenerate case it can reach a few MHz, which although smaller than the $XX$ coupling it is still significant and can introduce errors in qubits operations. Many more $ZZ$ couplings will appear as the number of qubits increases due to the many ways a qubit can interact with the other qubits when there is an all-to-all connection between them.

We studied the effect of qubit detuning on the qubits effective coupling by detuning two of the qubits up and down by the same amount $\Delta\omega$ respectively from the frequency of the third qubit as it was experimentally done for phase qubits. For the single-excitation case we did not calculate the $XX$ coupling but used the error state occupation probability, which showed a $\Delta\omega^{-2}$ decay with detuning when the later is large compared to the characteristic capacitive coupling strength. 

For the three pairwise and the only all-to-all $ZZ$ couplings, all four decay to zero for large detuning but before that they show spikes at detuning values corresponding to the resonances with nearly-degenerate qubit states out of the computational basis. The spikes in $ZZ$ coupling can become of the same order of magnitude as the $XX$ coupling and thus lead to significant qubit operation errors. The largest detuning values for which the $ZZ$ couplings spike are given by the anharmonicity of both the base (undetuned) qubit and the qubit which frequency was tuned up in frequency, which for typical system's parameters it is around 200 MHz. Thus, the qubits will need to be detuned well beyond this value in order to significantly suppress all the $ZZ$ couplings. 

It was proposed that having a system with tunable couplers could help controlling and diminishing the $ZZ$ coupling between qubits \cite{Neeley2010}, but this still has to be explored in more detail for a highly-connected system like the one we studied here, where we have shown that more than one type of $ZZ$ coupling exist.

\appendix

\section{Derivation of the system's Hamiltonian}\label{hamiltonian}

Using the Kirchhoff circuit laws we have the following equation for the voltages $V_{xi}$ across the coupling capacitors $C_{xi}$ \cite{Likharev1986, Barone1982, Tinkham1996, Krantz2019}:
\begin{eqnarray}
	-V_{x2} - V_2 + V_1 + V_{x1} &=& 0, \label{eq:k1} \\
	-V_{x3} - V_3 + V_3 + V_{x3} &=& 0, \label{eq:k2} \\
	-I_{x1} = I_{x2} + I_{x3}, \label{eq:curreq}
\end{eqnarray}
where $V_i$ is the voltage across the qubit with capacitance $C_i$ and the currents $I_{xi}$ are directed as shown in Fig.\ref{system}-b. Assuming that the initial charge at the island is zero, Eq. (\ref{eq:curreq}) implies that
\begin{equation}
	-C_{x1}V_{x1} - C_{x2}V_{x2} - C_{x3}V_{x3} = 0,
\end{equation}
which together with (\ref{eq:k1}) and (\ref{eq:k2}) allows us to write the voltages $V_{xi}$ in terms of the qubit voltages $V_i$:
\begin{eqnarray}
	V_{x1} &=& \frac{-1}{C_{\Sigma}}[-C_{x2}V_2 + (C_{x2}+C_{x3})V_1 - C_{x3}V_3 ],\\
	V_{x2} &=& \frac{-1}{C_{\Sigma}}[-C_{x3}V_3 + (C_{x3}+C_{x1})V_2 - C_{x1}V_1 ],\\
	V_{x3} &=& \frac{-1}{C_{\Sigma}}[-C_{x1}V_1 + (C_{x2}+C_{x1})V_3 - C_{x2}V_2 ],
\end{eqnarray}
where $C_{\Sigma} = C_{x1} + C_{x2} + C_{x3}$.
Using the relation $V_i = \Phi_0\dot{\varphi}_i/(2\pi)$, the Langrangian is
\begin{eqnarray}
	\mathcal{L} &=& \sum_{i=1}^{3}\frac{1}{2}C_iV_i^2 + \frac{1}{2}C_{xi}V_{xi}^2 + E_{J,i}\cos\varphi_i, \\
	&=& \sum_{i=1}^{3}\frac{1}{2}\left(\frac{\Phi_0}{2\pi}\right)^2 C_i\dot{\varphi}_i^2 {} \nonumber \\
	&& + \frac{1}{2}\left(\frac{\Phi_0}{2\pi}\right)^2\frac{C_{x1}}{C_{\Sigma}^2}\left[ \tilde{C}_{x1} \dot{\varphi}_1 - C_{x3}\dot{\varphi}_3 - C_{x2}\dot{\varphi_2} \right] {} \nonumber \\ 
	&& + \frac{1}{2}\left(\frac{\Phi_0}{2\pi}\right)^2\frac{C_{x2}}{C_{\Sigma}^2}\left[ \tilde{C}_{x2} \dot{\varphi}_1 - C_{x1}\dot{\varphi}_1 - C_{x3}\dot{\varphi_3} \right] {} \nonumber \\
	&& + \frac{1}{2}\left(\frac{\Phi_0}{2\pi}\right)^2\frac{C_{x3}}{C_{\Sigma}^2}\left[ \tilde{C}_{x3} \dot{\varphi}_3 - C_{x2}\dot{\varphi}_2 - C_{x1}\dot{\varphi_1} \right] {} \nonumber \\
	&& + E_{J,i}\cos\varphi_i,
\end{eqnarray}
\\
where $\tilde{C}_{xi} = \sum_{j\neq i}C_{xj}$ and $E_{J,i}$ are the qubit Josephson energies.

The conjugate momenta $p_i = \partial\mathcal{L}/\partial\dot{\varphi_i}$ are given by
\begin{equation}
	\mathbf{p} = \left(\frac{\Phi_0}{2\pi}\right)^2 \mathbf{C}\dot{\mathbf{\varphi}},
\end{equation}
where $\mathbf{p} = (p_1\;p_2\;p_3)^t$, $\mathbf{\varphi} = (\varphi_1\;\varphi_2\;\varphi_3)^t$, and $\mathbf{C}$ is the $3\times3$ capacitancce matrix of the system, with matrix elements
\begin{eqnarray}
	C_{ii} &=& C_i + \frac{C_{xi}}{C_{\Sigma}}\tilde{C}_{xi},\\
	C_{ij} &=& \frac{C_{xi}C_{xj}}{C_{\Sigma}}, \, j\neq i.
\end{eqnarray}
The Hamiltonian of the system is then
\begin{eqnarray}
	H = \sum_{i=1}^{3} \frac{1}{2} \left(\frac{\Phi_0}{2\pi}\right)^2 \left[\mathbf{C^{-1}}\right]_{ii} p_i^2 - E_{J,i}\cos\varphi_i {} \nonumber \\
	+ \sum_{i=1}^{3}\sum_{j>i}^{3}  \left(\frac{\Phi_0}{2\pi}\right)^2 \left[\mathbf{C^{-1}}\right]_{ij} p_ip_j.
\end{eqnarray}

Introducing the shifted variables $\delta\varphi_i = \varphi_i - \varphi_{i,st}$, where the
set $\{\varphi_{i,st}\}$ corresponds to the minimum of the potential energies $U_i(\delta\varphi_i) = -\left[E_{J,i}\cos(\delta\varphi_i + \varphi_{i,st}) - E_{J,i}\cos(\varphi_{i,st})\right]$ and defines the qubit plasma frequencies  $\omega_{pl,i} = \sqrt{E_{J,i}\cos(\varphi_{i,st})/m_{i}}$, we can rewrite the Hamiltonian as
\begin{equation}
	H = \sum_{i=1}^{3} H_i + H_{int},
\end{equation}
where
\begin{eqnarray}
	H_i &=& \frac{1}{2} \left(\frac{\Phi_0}{2\pi}\right)^2 \left[\mathbf{C^{-1}}\right]_{ii} p_i^2 + U_i(\delta\varphi_i), \\
	H_{int} &=&  \sum_{i=1}^{3}\sum_{j>i}^{3}  \left(\frac{\Phi_0}{2\pi}\right)^2 \left[\mathbf{C^{-1}}\right]_{ij} p_ip_j.
\end{eqnarray}
and the potentials $U_i$ have minima at $\delta\varphi_i = 0$.

\section{Dressed states from first-order perturbation theory}\label{dressed}

To lowest order in perturbation theory, and neglecting terms with four or more total qubits excitation number, we have
\begin{eqnarray}
	|\psi_{100}^{dr}\rangle &=& |100\rangle - \frac{\sqrt{2}K_{12}}{E - \epsilon_{210}}|210\rangle - \frac{\sqrt{2}K_{13}}{E - \epsilon_{201}}|201\rangle {} \nonumber \\ 
	& & - \frac{K_{23}}{E - \epsilon_{111}}|111\rangle ,\\
	|\psi_{010}^{dr}\rangle &=& |010\rangle - \frac{K_{12}\sqrt{2}}{E - \epsilon_{120}}|120\rangle - \frac{K_{23}\sqrt{2}}{E - \epsilon_{021}}|021\rangle {} \nonumber \\ 
	& & - \frac{K_{13}}{E - \epsilon_{111}}|111\rangle ,\\
	|\psi_{001}^{dr}\rangle &=& |001\rangle - \frac{\sqrt{2}K_{13}}{E - \epsilon_{102}}|102\rangle - \frac{\sqrt{2}K_{23}}{E - \epsilon_{012}}|012\rangle {} \nonumber \\ 
	& & - \frac{K_{12}}{E - \epsilon_{111}}|111\rangle ,
\end{eqnarray}
where $\epsilon_{n_1n_2n_3}$ are the eigenenergies of the non-interacting system. The corresponding self-energies are
\begin{eqnarray}
	E_{100}^{dr} &=& \epsilon_{100} + \frac{2K_{12}^2}{E - \epsilon_{210}} + \frac{2K_{13}^2}{E - \epsilon_{201}} + \frac{K_{23}^2}{E - \epsilon_{111}}, \\	
	E_{010}^{dr} &=& \epsilon_{010} + \frac{2K_{12}^2}{E - \epsilon_{120}} + \frac{K_{13}^2}{E - \epsilon_{111}} + \frac{2K_{23}^2}{E - \epsilon_{021}}, \\	
	E_{001}^{dr} &=& \epsilon_{001} + \frac{K_{12}^2}{E - \epsilon_{111}} + \frac{2K_{13}^2}{E - \epsilon_{102}} + \frac{2K_{23}^2}{E - \epsilon_{012}};
\end{eqnarray}
and the effective interactions are
\begin{eqnarray}
	V_{12}^{dr} &=& V_{21}^{dr} = K_{12} + \frac{K_{13}K_{23}}{E - \epsilon_{111}}, \\
	V_{13}^{dr} &=& V_{31}^{dr} = K_{13} + \frac{K_{12}K_{23}}{E - \epsilon_{111}}, \\
	V_{23}^{dr} &=& V_{32}^{dr} = K_{23} + \frac{K_{13}K_{12}}{E - \epsilon_{111}}.
\end{eqnarray}
For weak coupling ($C_{xi} \ll C_i$, $i=1,\,2,\,3$)
\begin{eqnarray}
	K_{12} &\simeq& -\frac{C_{x1}C_{x2}}{\sqrt{C_{11}C_{22}}C_{\Sigma}} \frac{\hbar\sqrt{\omega_1\omega_2}}{2}, \\
	K_{13} &\simeq& -\frac{C_{x1}C_{x3}}{\sqrt{C_{11}C_{33}}C_{\Sigma}} \frac{\hbar\sqrt{\omega_1\omega_3}}{2}, \\
	K_{23} &\simeq& -\frac{C_{x2}C_{x3}}{\sqrt{C_{22}C_{33}}C_{\Sigma}} \frac{\hbar\sqrt{\omega_2\omega_3}}{2};
\end{eqnarray}
and we can approximate $E \approx \epsilon_{100},\, \epsilon_{001},\, \epsilon_{001}$ in each of the corresponding dressed states and couplings. For degenerate qubits ($\omega_i = \omega_{qb}$) this leads to $E - \epsilon_{210} \approx E - \epsilon_{201} \approx E - \epsilon_{111} \approx -2\hbar\omega_{qb}$ and hence to:
\begin{eqnarray}
	E_{100}^{dr} &\approx& (1 - \xi) \hbar\omega_{qb}\\
	V_{12}^{dr} &\approx& K_{12} \simeq -\frac{C_{x1}C_{x2}}{\sqrt{C_{11}C_{22}}C_{\Sigma}} \frac{\hbar\omega_{qb}}{2} \label{eq:Vdr},	
\end{eqnarray}
where $\xi = \frac{\left( C_{x1}C_{x2} \right)^2}{C_{11}C_{22}C_{\Sigma}^2}
+ \frac{\left( C_{x1}C_{x3} \right)^2}{C_{11}C_{33}C_{\Sigma}^2} + \frac{\left( C_{x2}C_{x3} \right)^2}{C_{22}C_{33}C_{\Sigma}^2}$. 

\section{Quantum couplings --degenerate qubits}\label{couplingdegeneratederivation}

\subsection{Pairwise $ZZ$ coupling $\zeta_p$}

For the case of having a completely symmetric system ($C_i = C$ and $C_{xi} = C_x$ for $i=1,\,2,\,3$), we have $E_{100}^{dr} = E_{010}^{dr} = E_{001}^{dr} \equiv E_{[100]}^{dr}$ and $V_{12}^{dr} = V_{13}^{dr} = V_{23}^{dr} \equiv V^{dr}$, and Eqs.(\ref{eq:alpha}-\ref{eq:gamma}) become
\begin{eqnarray}
	E_{[100]}^{dr}\alpha + V^{dr}\beta + V^{dr}\gamma &=& E\alpha,\label{eq:eigen1} \\
	V^{dr}\alpha + E_{[100]}^{dr}\beta + V^{dr}\gamma &=& E\beta, \label{eq:eigen2} \\	
	V^{dr}\alpha + V^{dr}\beta + E_{[100]}^{dr}\gamma &=& E\gamma \label{eq:eigen3},	
\end{eqnarray} 
where
\begin{eqnarray}
	E_{[100]}^{dr} &\approx& \left[ 1 - \frac{5}{8} \left(\frac{C_x}{3C_D}\right)^2 \right] \hbar\omega_{qb}, \label{eq:E100dr}\\
	V^{dr} &\approx& - \frac{C_x}{3C_D} \frac{\hbar\omega_{qb}}{2},
\end{eqnarray}
and $C_{11} = C_{22} = C_{33} \equiv C_D = C + 2C_x/3 \simeq C$.

Solution of the eigenvalue problem above leads to the three eigenvalues $E_1 = E_{[100]}^{dr} + 2V^{dr} \equiv E_-$ and $E_2 = E_3 = E_{[100]}^{dr} - V^{dr} \equiv E_+$, and to the $XX$ coupling in Eq. (\ref{eq:OmegaXX}).

To derive the pairwise $ZZ$ coupling $\zeta_p$ one can carry out a similar analysis and derive the eigenenergies $E_{110},\,E_{101},\,E_{011}$ by constructing the dressed states $|\psi_{110}^{dr}\rangle,\,|\psi_{101}^{dr}\rangle$, and $|\psi_{011}^{dr}\rangle$ which for the degenerate case leads to equations like Eqs. (\ref{eq:eigen1}-\ref{eq:eigen3}) except that the self energies are now $E_{[110]}^{dr}$. Thus the eigenenergies $E_{110},\,E_{101},\,E_{011}$ correspond to the eigenvalues  $E_4 = E_{[110]}^{dr} + 2V^{dr}$ and $E_5 = E_6 = E_{[110]}^{dr} - V^{dr}$. 

To lowest order in perturbation theory, the self-energy $E_{110}^{dr}$ is
\begin{eqnarray}
	E_{110}^{dr} \simeq & & \epsilon_{110} + K_{12}^2 \left(\frac{1}{\epsilon_{110} - \epsilon_{000}} + \frac{4}{\epsilon_{110} - \epsilon_{220}}\right) \nonumber \\
	& & + \frac{\epsilon_{200} - \epsilon_{110} \pm \sqrt{(\epsilon_{200} - \epsilon_{110})^2 + S_{|110\rangle,|200\rangle}^2}}{2} \nonumber \\
	& & + \frac{\epsilon_{020} - \epsilon_{110} \pm \sqrt{(\epsilon_{020} - \epsilon_{110})^2 + S_{|110\rangle,|020\rangle}^2}}{2} \label{eq:E110dr},
\end{eqnarray}
where the last two terms are the corrections due to the avoided level crossing interaction between state $|110\rangle$ and nearly degenerate states $|200\rangle$ and $|020\rangle$ respectively with effective couplings $S_{|110\rangle,|200\rangle} = 2\langle110|H|\psi_{200}^{dr}\rangle$ and $S_{|110\rangle,|020\rangle} = 2\langle110|H|\psi_{020}^{dr}\rangle$.

The dressed state $|\psi_{200}^{dr}\rangle$ is constructed as satisfying the Schr\"odinger equation $\langle n_1n_2n_3|H|\psi_{200}^{dr}\rangle = E\langle n_1n_2n_3|\psi_{200}^{dr}\rangle$ for all basis elements $|n_1n_2n_3\rangle$ except $|110\rangle$, $|101\rangle$, $|011\rangle$, $|020\rangle$, and $|002\rangle$; assuming $E \approx \epsilon_{200}$. Similarly, the dressed state $|\psi_{020}^{dr}\rangle$ is constructed as satisfying the Schr\"odinger equation $\langle n_1n_2n_3|H|\psi_{020}^{dr}\rangle = E\langle n_1n_2n_3|\psi_{020}^{dr}\rangle$ for all basis elements $|n_1n_2n_3\rangle$ except $|110\rangle$, $|101\rangle$, $|011\rangle$, $|200\rangle$, and $|002\rangle$; assuming $E \approx \epsilon_{020}$. Finally, the dressed state $|\psi_{002}^{dr}\rangle$ is constructed as satisfying the Schr\"odinger equation $\langle n_1n_2n_3|H|\psi_{002}^{dr}\rangle = E\langle n_1n_2n_3|\psi_{002}^{dr}\rangle$ for all basis elements $|n_1n_2n_3\rangle$ except $|110\rangle$, $|101\rangle$, $|011\rangle$, $|200\rangle$, and $|020\rangle$; assuming $E \approx \epsilon_{002}$. To lowest order in perturbation theory within the harmonic approximation, they are:
\begin{eqnarray}
	|\psi_{200}^{dr}\rangle &=& |200\rangle - \frac{\sqrt{3}K_{12}}{E - \epsilon_{310}}|310\rangle - \frac{\sqrt{3}K_{13}}{E - \epsilon_{301}}|301\rangle {} \nonumber \\ 
	& & - \frac{K_{23}}{E - \epsilon_{211}}|211\rangle \label{eq:psi200},\\
	|\psi_{020}^{dr}\rangle &=& |020\rangle - \frac{\sqrt{3}K_{23}}{E - \epsilon_{031}}|031\rangle - \frac{\sqrt{3}K_{12}}{E - \epsilon_{130}}|130\rangle {} \nonumber \\ 
	& & - \frac{K_{13}}{E - \epsilon_{121}}|121\rangle \label{eq:psi020},\\
	|\psi_{002}^{dr}\rangle &=& |002\rangle - \frac{\sqrt{3}K_{13}}{E - \epsilon_{103}}|103\rangle - \frac{\sqrt{3}K_{23}}{E - \epsilon_{013}}|013\rangle {} \nonumber \\ 
	& & - \frac{K_{12}}{E - \epsilon_{112}}|112\rangle \label{eq:psi002}.
\end{eqnarray}
Thus, $S_{|110\rangle,|200\rangle} \simeq 2\sqrt{2}K_{12}$ and $S_{|110\rangle,|020\rangle} \simeq 2\sqrt{2}K_{12}$, and Eq.(\ref{eq:E110dr}) becomes for the degenerate case where $K_{12} = K = -[C_x/(3C)]\hbar\omega_{qb}/2$:
\begin{eqnarray}
	E_{[110]}^{dr} \simeq & & 2\hbar\omega_{qb} + \left(\frac{C_x}{3C}\right)^2\frac{\hbar\omega_{qb}}{8} \left(1 - \frac{4}{1 + \alpha_{qb}}\right) \nonumber \\
	& & + \hbar\omega_{qb}\left[\sqrt{\alpha_{qb}^2 + 2\left(\frac{C_x}{3C}\right)^2} + \alpha_{qb} \right],
\end{eqnarray}
where $\alpha_{qb} = (\omega_{12}-\omega_{qb})/\omega_{qb}$ is the relative anharmonicity and we take the positive square root in Eq.(\ref{eq:E110dr}) because $\epsilon_{110} > \epsilon_{200}, \epsilon_{020}$ and thus $E_{110}$ is being pushed up in the avoided level crossing interaction with states $|200\rangle$ and $|020\rangle$.

The self energy $E_{000}^{dr}$, to lowest order in perturbation theory, is
\begin{equation}
	E_{000}^{dr} \simeq \epsilon_{000} + \frac{K_{12}^2}{\epsilon_{000} - \epsilon_{110}} + \frac{K_{13}^2}{\epsilon_{000} - \epsilon_{101}} + \frac{K_{23}^2}{\epsilon_{000} - \epsilon_{011}},
\end{equation}
where $\epsilon_{000} = 0$. In the degenerate case the equation above reduces to
\begin{equation}
	E_{000}^{dr} \simeq -\frac{3}{2}\frac{K^2}{\hbar\omega_{qb}} = -\frac{3}{8}\left(\frac{C_x}{3C}\right)^2\hbar\omega_{qb},
\end{equation}
In Eq. (\ref{eq:ZZp}), $E_{100}+E_{010}+E_{001} = E_1+E_2+E_3 = E_- + 2E_+ = 3E_{[100]}^{dr}$, and $E_{110}+E_{101}+E_{011} = E_7+E_8+E_9 = 3E_{[110]}^{dr}$. With the self-energy $E_{[100]}^{dr}$ given by Eq.(\ref{eq:E100dr}) we obtain the pairwise $ZZ$ coupling $\zeta_p$ in Eq.(\ref{eq:zzp}).

\subsection{Three-qubits (all-to-all) $ZZ$ coupling $\zeta_{111}^{qb}$}

To calculate the three-qubits $ZZ$ coupling for the degenerate symmetric case, $\zeta_{111}^{qb}$, we construct the dressed state $|\psi_{111}^{dr}\rangle$ as satisfying the Schr\"odinger equation $\langle n_1n_2n_3|H|\psi_{111}^{dr}\rangle = E\langle n_1n_2n_3|\psi_{111}^{dr}\rangle$ for all basis elements $|n_1n_2n_3\rangle$ except $|111\rangle$. To lowest order in perturbation theory, with $E \approx \epsilon_{111}$,
\begin{eqnarray}
	|\psi_{111}^{dr}\rangle &=& |111\rangle - K_{12} \Bigg( \frac{|001\rangle}{E - \epsilon_{001}}  + \frac{2|221\rangle}{E - \epsilon_{221}} - \frac{\sqrt{2}|201\rangle}{E - \epsilon_{201}} \nonumber \\
	& & - \frac{\sqrt{2}|021\rangle}{E - \epsilon_{021}} \Bigg)
	- K_{13} \Bigg( \frac{|010\rangle}{E - \epsilon_{010}}  + \frac{2|212\rangle}{E - \epsilon_{212}} \nonumber \\
	& &  - \frac{\sqrt{2}|210\rangle}{E - \epsilon_{210}} -  \frac{\sqrt{2}|012\rangle}{E - \epsilon_{012}} \Bigg)
	- K_{23} \Bigg( \frac{|100\rangle}{E - \epsilon_{100}}  \nonumber \\ 
	& & + \frac{2|122\rangle}{E - \epsilon_{122}} - \frac{\sqrt{2}|120\rangle}{E - \epsilon_{120}} - \frac{\sqrt{2}|102\rangle}{E - \epsilon_{102}}  \Bigg),
\end{eqnarray}
and the self-energy $E_{111}^{dr}$ is
\begin{eqnarray}
	E_{111}^{dr} \simeq & & \epsilon_{111} + K_{12}^2 \left(\frac{1}{\epsilon_{111} - \epsilon_{001}} + \frac{2}{\epsilon_{111} - \epsilon_{201}} + \frac{2}{\epsilon_{11} - \epsilon_{021}} \right) \nonumber \\
	& & +  K_{13}^2 \left(\frac{1}{\epsilon_{111} - \epsilon_{010}} + \frac{2}{\epsilon_{111} - \epsilon_{210}} + + \frac{2}{\epsilon_{111} - \epsilon_{012}} \right) \nonumber \\
	& & +  K_{23}^2 \left(\frac{1}{\epsilon_{111} - \epsilon_{100}} + \frac{2}{\epsilon_{111} - \epsilon_{120}} + \frac{2}{\epsilon_{111} - \epsilon_{102}} \right). \nonumber \\ 
\end{eqnarray}
For degenerate qubits $E_{111}^{dr}$ becomes
\begin{equation}
	E_{111}^{dr} \simeq \epsilon_{111} + \frac{3K^2}{\hbar\omega_{qb}} \left(\frac{4}{\alpha_{qb}} - \frac{3}{2}\right),
\end{equation}
and substitution into Eq.(\ref{eq:zz111}) leads to Eq.(\ref{eq:ZZ111qb}).

\section{Single-excitation eigenstates and error state occupation --detuned qubits}\label{errorstatederivation}

A single-excitation eigenstate $|\psi\rangle$ of the system can be approximate as
\begin{equation}
	|\psi\rangle = \alpha|\psi_{100}^{dr}\rangle + \beta|\psi_{010}^{dr}\rangle + \gamma|\psi_{001}^{dr}\rangle.
\end{equation}
The coefficients $\alpha,\,\beta$, and $\gamma$ satisfy the system of equations (\ref{eq:alpha}-\ref{eq:gamma}) that we rewrite here:
\begin{eqnarray}
	E_{100}^{dr}\alpha + V_{12}^{dr}\beta + V_{13}^{dr}\gamma &=& E\alpha, \label{eq:app_alpha}\\
	V_{21}^{dr}\alpha + E_{010}^{dr}\beta + V_{23}\gamma &=& E\beta, \label{eq:app_beta}\\	
	V_{31}\alpha + V_{32}\beta + E_{001}^{dr}\gamma &=& E\gamma \label{eq:app_gamma},
\end{eqnarray} 
where $E_{100}^{dr} \equiv \langle 100|H|\psi_{100}^{dr}\rangle$, $E_{010}^{dr} \equiv \langle 010|H|\psi_{010}^{dr}\rangle$, and $E_{001}^{dr} \equiv \langle 001|H|\psi_{001}^{dr}\rangle$; and $V_{12}^{dr} \equiv \langle 100|H|\psi_{010}^{dr}\rangle$, $V_{21}^{dr} \equiv \langle 010|H|\psi_{100}^{dr}\rangle$, $V_{13}^{dr} \equiv \langle 100|H|\psi_{001}^{dr}\rangle$, $V_{31}^{dr} \equiv \langle 001|H|\psi_{100}^{dr}\rangle$, $V_{23}^{dr} \equiv \langle 010|H|\psi_{001}^{dr}\rangle$, and $V_{32}^{dr} \equiv \langle 001|H|\psi_{010}^{dr}\rangle$.

For weak qubit coupling $V_{ij}^{dr} \ll \hbar\omega_i$, $E_{100}^{dr} \simeq \epsilon_{100} = \hbar\omega_1 = \hbar\omega_2 - \Delta\omega$, $E_{010}^{dr} \simeq \epsilon_{010} = \hbar\omega_2$, and $E_{001}^{dr} \simeq \epsilon_{001} = \hbar\omega_3 =  \hbar\omega_2 + \Delta\omega$; and in the symmetric case the effective interactions $V_{ij}^{dr}$ become
\begin{eqnarray}
	V_{12}^{dr} &\approx& K_{12} = K\sqrt{1-\frac{\Delta\omega}{\omega_2}}, \\
	V_{13}^{dr} &\approx& K_{13} = K\sqrt{1+\frac{\Delta\omega}{\omega_2}}, \\
	V_{23}^{dr} &\approx& K_{23} = K\sqrt{1-\left(\frac{\Delta\omega}{\omega_2}\right)^2},
\end{eqnarray}
where
\begin{equation}
	K = -\frac{C_x}{3C}\frac{\hbar\omega_2}{2}.
\end{equation}

From Eqs.(\ref{eq:app_beta}-\ref{eq:app_gamma}) one obtains
\begin{eqnarray}
	\beta &=& \frac{\left[\left(E_{001}^{dr}-E\right)V_{12}^{dr} - V_{12}^{dr}V_{23}^{dr}\right]}{(V_{23}^{dr})^2 - \left(E_{010}^{dr} - E\right)\left(E_{001}^{dr} - E\right)} \alpha, \label{eq:app_beta2}\\	
	\gamma &=& \frac{\left[\left(E_{100}^{dr}-E\right)V_{13}^{dr} - V_{12}^{dr}V_{23}^{dr}\right]}{(V_{23}^{dr})^2 - \left(E_{010}^{dr} - E\right)\left(E_{001}^{dr} - E\right)} \alpha \label{eq:app_gamma2}.
\end{eqnarray}

We also assume large qubit detuning, $\Delta\omega \gg |K|/\hbar$, where the eigenenergies are close to the system's energies in the non-interacting case, $E = E_{100} \simeq \epsilon_{100} = \hbar(\omega_2 - \Delta\omega)$, $E = E_{010} \simeq \epsilon_{010} = \hbar\omega_2$, $E = E_{001} \simeq \epsilon_{001} = \hbar(\omega_2 + \Delta\omega)$. Using these eigenenergies in Eqs.(\ref{eq:app_beta2}-\ref{eq:app_gamma2}) for $\alpha=1$, one obtains the corresponding normalized eigenstates given in Eqs.(\ref{eq:psi100detuned}-\ref{eq:psi001detuned}).

\section{ZZ coupling --detuned qubits}\label{zzdetunedderivation}

\subsection{Pairwise $ZZ$ coupling $\zeta_{110}$}

For large qubit detuning $\Delta\omega \gg |K|/\hbar$, the correction to $E_{110} \simeq \epsilon_{110} = \hbar(2\omega_{2}-\Delta\omega)$ due to interaction with state $|200\rangle$ is $[\epsilon_{200}-\epsilon_{110} \pm \sqrt{(\epsilon_{200}-\epsilon_{110})^2 + S_{|110\rangle,|200\rangle}^2}]/2$, where $S_{|110\rangle,|200\rangle} = 2\langle 110|H|\psi_{200}^{dr}\rangle$. 
Similarly, the corrections to $E_{110}$ due to interaction with states $|020\rangle$ and $|002\rangle$ are $[\epsilon_{020}-\epsilon_{110} \pm \sqrt{(\epsilon_{020}-\epsilon_{110})^2 + S_{|110\rangle,|020\rangle}^2}]/2$ and $[\epsilon_{002}-\epsilon_{110} \pm \sqrt{(\epsilon_{002}-\epsilon_{110})^2 + S_{|110\rangle,|002\rangle}^2}]/2$ respectively.

With $|\psi_{200}^{dr}\rangle$, $|\psi_{020}^{dr}\rangle$, and $|\psi_{002}^{dr}\rangle$ given by (\ref{eq:psi200}-\ref{eq:psi002}) one obtains $S_{|110\rangle,|200\rangle} \approx 2\sqrt{2}K_{12}$ and $S_{|110\rangle,|020\rangle} \approx 2\sqrt{2}K_{12}$, while $S_{|110\rangle,|002\rangle}  = 2\langle 110|H|\psi_{002}^{dr}\rangle$ is a higher-order effective interaction mediated by the states $|101\rangle$ and $|011\rangle$: 
\begin{equation}
	S_{|110\rangle,|002\rangle} \approx 2\sqrt{2}K_{13}K_{23} \left(\frac{1}{\varepsilon_{1+} - \varepsilon_{1-}} + \frac{1}{\varepsilon_{1+}^{\prime} - \varepsilon_{1-}^{\prime}}\right);
\end{equation}
where the energies
\begin{equation}
	\varepsilon_{1\pm} = \frac{1}{2}\left[ \epsilon_{002}+\epsilon_{101} \pm \sqrt{(\epsilon_{002}-\epsilon_{101})^2 + S_{|101\rangle,|002\rangle}^2} \right]
\end{equation}
replace the nearly degenerate energies $\epsilon_{002}$ and $\epsilon_{101}$, with $S_{|101\rangle,|002\rangle} = 2\langle 101|H|\psi_{002}^{dr}\rangle \approx 2\sqrt{2}K_{13}$; and the energies
\begin{equation}
	\varepsilon_{1\pm}^{\prime} = \frac{1}{2}\left[ \epsilon_{002}+\epsilon_{011} \pm \sqrt{(\epsilon_{002}-\epsilon_{011})^2 + S_{|011\rangle,|002\rangle}^2} \right]
\end{equation}
replace $\epsilon_{002}$ and $\epsilon_{011}$, with $S_{|011\rangle,|002\rangle} = 2\langle 011|H|\psi_{002}^{dr}\rangle \approx 2\sqrt{2}K_{23}$.

The double-excitation energies are $\epsilon_{200} = \hbar\omega_1(2-\alpha_1) = \hbar(\omega_2-\Delta\omega)(2-\alpha_1)$, $\epsilon_{020} = \hbar\omega_2(2-\alpha_2)$, and $\epsilon_{002} = \hbar\omega_3(2-\alpha_3) = \hbar(\omega_2+\Delta\omega)(2-\alpha_1)$, where $\alpha_{i} = [\omega_{12}^{(i)}-\omega_i]/\omega_i$ and $\omega_{12}^{(i)}$ are the relative anharmonicity and $|1\rangle-|2\rangle$ transition frequency of qubit $i$ respectively.

Combining $E_{110}$ with $E_{000} \simeq \epsilon_{000} = 0$, $E_{100} \simeq \epsilon_{100} = \hbar(\omega_2 - \Delta\omega)$, $E_{010} \simeq \epsilon_{010} = \hbar\omega_2$, and $E_{001} \simeq \epsilon_{001} = \hbar(\omega_2 + \Delta\omega)$ in Eq.(\ref{eq:zz110}) one obtains (\ref{eq:zeta110}) for $\zeta_{110}$.

\subsection{Pairwise $ZZ$ coupling $\zeta_{101}$}

For large qubit detuning $\Delta\omega \gg |K|/\hbar$, the correction to $E_{101} \simeq \epsilon_{101} = 2\hbar\omega_{2}$ due to interaction with state $|200\rangle$ is $[\epsilon_{200}-\epsilon_{101} \pm \sqrt{(\epsilon_{200}-\epsilon_{101})^2 + S_{|101\rangle,|200\rangle}^2}]/2$, where $S_{|101\rangle,|200\rangle} = 2\langle 101|H|\psi_{200}^{dr}\rangle$. 
Similarly, the corrections to $E_{101}$ due to interaction with states $|002\rangle$ and $|020\rangle$ are $[\epsilon_{002}-\epsilon_{101} \pm \sqrt{(\epsilon_{002}-\epsilon_{101})^2 + S_{|101\rangle,|002\rangle}^2}]/2$ and $[\epsilon_{020}-\epsilon_{101} \pm \sqrt{(\epsilon_{020}-\epsilon_{101})^2 + S_{|101\rangle,|020\rangle}^2}]/2$ respectively.

One obtains $S_{|101\rangle,|200\rangle} \approx 2\sqrt{2}K_{13}$ and $S_{|101\rangle,|002\rangle} \approx 2\sqrt{2}K_{13}$, while $S_{|101\rangle,|020\rangle}  = 2\langle 101|H|\psi_{020}^{dr}\rangle$ is a higher-order effective interaction mediated by the states $|110\rangle$ and $|011\rangle$: 
\begin{equation}
	S_{|101\rangle,|020\rangle} \approx 2\sqrt{2}K_{13}K_{23} \left(\frac{1}{\varepsilon_{2+} - \varepsilon_{2-}} + \frac{1}{\varepsilon_{2+}^{\prime} - \varepsilon_{2-}^{\prime}}\right); \label{eq:S101_002}
\end{equation}
where the energies
\begin{equation}
	\varepsilon_{2\pm} = \frac{1}{2}\left[ \epsilon_{020}+\epsilon_{110} \pm \sqrt{(\epsilon_{020}-\epsilon_{110})^2 + S_{|110\rangle,|020\rangle}^2} \right]
\end{equation}
replace the nearly degenerate energies $\epsilon_{020}$ and $\epsilon_{110}$, with $S_{|110\rangle,|020\rangle} = 2\langle 110|H|\psi_{020}^{dr}\rangle \approx 2\sqrt{2}K_{12}$; and the energies
\begin{equation}
	\varepsilon_{2\pm}^{\prime} = \frac{1}{2}\left[ \epsilon_{020}+\epsilon_{011} \pm \sqrt{(\epsilon_{020}-\epsilon_{011})^2 + S_{|011\rangle,|020\rangle}^2} \right]
\end{equation}
replace $\epsilon_{020}$ and $\epsilon_{011}$, with $S_{|011\rangle,|020\rangle} = 2\langle 011|H|\psi_{020}^{dr}\rangle \approx 2\sqrt{2}K_{23}$.

The double-excitation energies $\epsilon_{200}$, $\epsilon_{020}$, and $\epsilon_{002}$ in terms in the qubit anharmonicities were already given at the end of the previous subsection; and combining $E_{101}$ with $E_{000}$, $E_{100}$, and $E_{001}$ in Eq.(\ref{eq:zz101}) one obtains the pairwise $ZZ$ coupling $\zeta_{101}$:
\begin{eqnarray}
	\zeta_{101}(\Delta\omega) \simeq && \frac{\omega_2}{2} \left[\alpha_1 + \alpha_3 + (\alpha_3-\alpha_1)\frac{\Delta\omega}{\omega_2} \right]   \nonumber \\
	&& + \frac{\omega_2}{2} \mathcal{P}\left[\frac{(2+\alpha_1)\Delta\omega}{\omega_2} - \alpha_1; \frac{2K_{13}}{\hbar\omega_2}\right] \nonumber \\
	&& - \frac{\omega_2}{2} \mathcal{P}\left[\frac{(2+\alpha_3)\Delta\omega}{\omega_2}+\alpha_3; \frac{2K_{13}}{\hbar\omega_2}\right], \label{eq:zeta101}
\end{eqnarray}
where
\begin{equation}
	\mathcal{P}(x; k) = \left[2\theta(x) - 1\right] \sqrt{ x^2 + 2k^2 },
\end{equation}
$\theta(x)$ is the Heaviside function.

\subsection{Pairwise $ZZ$ coupling $\zeta_{011}$}

For large qubit detuning $\Delta\omega \gg |K|/\hbar$, the correction to $E_{011} \simeq \epsilon_{011} = \hbar(2\omega_{2}+\Delta\omega$ due to interaction with state $|002\rangle$ is $[\epsilon_{002}-\epsilon_{011} \pm \sqrt{(\epsilon_{002}-\epsilon_{011})^2 + S_{|011\rangle,|002\rangle}^2}]/2$, where $S_{|011\rangle,|002\rangle} = 2\langle 011|H|\psi_{002}^{dr}\rangle$. 
Similarly, the corrections to $E_{011}$ due to interaction with states $|020\rangle$ and $|200\rangle$ are $[\epsilon_{020}-\epsilon_{011} \pm \sqrt{(\epsilon_{020}-\epsilon_{011})^2 + S_{|011\rangle,|020\rangle}^2}]/2$ and $[\epsilon_{200}-\epsilon_{011} \pm \sqrt{(\epsilon_{200}-\epsilon_{011})^2 + S_{|011\rangle,|200\rangle}^2}]/2$ respectively.

One obtains $S_{|011\rangle,|002\rangle} \approx 2\sqrt{2}K_{23}$ and $S_{|011\rangle,|020\rangle} \approx 2\sqrt{2}K_{23}$, while $S_{|011\rangle,|200\rangle}  = 2\langle 011|H|\psi_{200}^{dr}\rangle$ is a higher-order effective interaction mediated by the states $|110\rangle$ and $|101\rangle$: 
\begin{equation}
	S_{|011\rangle,|200\rangle} \approx 2\sqrt{2}K_{12}K_{13} \left(\frac{1}{\varepsilon_{3+} - \varepsilon_{3-}} + \frac{1}{\varepsilon_{3+}^{\prime} - \varepsilon_{3-}^{\prime}}\right); \label{eq:S011_200}
\end{equation}
where the energies
\begin{equation}
	\varepsilon_{3\pm} = \frac{1}{2}\left[ \epsilon_{200}+\epsilon_{110} \pm \sqrt{(\epsilon_{200}-\epsilon_{110})^2 + S_{|110\rangle,|200\rangle}^2} \right]
\end{equation}
replace the nearly degenerate energies $\epsilon_{200}$ and $\epsilon_{110}$, with $S_{|110\rangle,|200\rangle} = 2\langle 110|H|\psi_{200}^{dr}\rangle \approx 2\sqrt{2}K_{12}$; and the energies
\begin{equation}
	\varepsilon_{3\pm}^{\prime} = \frac{1}{2}\left[ \epsilon_{200}+\epsilon_{101} \pm \sqrt{(\epsilon_{200}-\epsilon_{101})^2 + S_{|101\rangle,|200\rangle}^2} \right]
\end{equation}
replace $\epsilon_{200}$ and $\epsilon_{101}$, with $S_{|101\rangle,|200\rangle} = 2\langle 101|H|\psi_{200}^{dr}\rangle \approx 2\sqrt{2}K_{13}$.

Combining $E_{011}$ with $E_{000}$, $E_{010}$, and $E_{001}$ in Eq.(\ref{eq:zz011}) one obtains the pairwise $ZZ$ coupling $\zeta_{011}$:
\begin{eqnarray}
	\zeta_{011}(\Delta\omega) \simeq && \frac{\omega_2}{2} \left[\alpha_1+\alpha_2+\alpha_3 + \left(1-\alpha_3\right)\frac{\Delta\omega}{\omega_2} \right]   \nonumber \\
	&& + \frac{\omega_2}{2} \mathcal{P}\left[\frac{\Delta\omega}{\omega_2} - \alpha_2; \frac{2K_{23}}{\hbar\omega_2}\right]   \nonumber \\
	&& - \frac{\omega_2}{2} \mathcal{P}\left[\frac{(1+\alpha_3)\Delta\omega}{\omega_2}+\alpha_3; \frac{2K_{23}}{\hbar\omega_2}\right]\nonumber \\
	&& + \frac{\omega_2}{2} \mathcal{P}\left[\frac{\Delta\omega}{\omega_2} - \alpha_1; \frac{\mathcal{S}_{|011\rangle,|200\rangle}}{\sqrt{2}}\right], \label{eq:zeta011}
\end{eqnarray}
where
\begin{widetext}
	\begin{equation}
		\mathcal{S}_{|011\rangle, |200\rangle} = \left(\frac{C_x}{3C}\right)^2 \frac{\sqrt{1-\frac{\Delta\omega}{\omega_2}}}{\sqrt{2}} \left\{ \frac{1}{\sqrt{\left[\frac{(2+\alpha_1)\Delta\omega}{\omega_2}-\alpha_1\right]^2 + 2\left(\frac{C_x}{3C}\right)^2}} 
		+ \frac{1}{\sqrt{\left[\frac{(1+\alpha_1)\Delta\omega}{\omega_2}-\alpha_1\right]^2 + 2\left(\frac{C_x}{3C}\right)^2\left(1-\frac{\Delta\omega}{\omega_2}\right) } } \right\}.
	\end{equation}
\end{widetext}

\subsection{Three-qubits (all-to-all) $ZZ$ coupling $\zeta_{111}$}

The eigenenergy $E_{111}$ approximates the bare energy $\epsilon_{111}$ and we have to add the corrections due to resonant interactions of the state $|111\rangle$ with states $|210\rangle,\,|021\rangle,\,|102\rangle,\,|201\rangle,\,|120\rangle$, and $|012\rangle$ (interaction with states $|300\rangle,\,|030\rangle$, and $|003\rangle$ are of higher order and thus their contribution is negligible).

For large qubit detuning $\Delta\omega \gg |K|/\hbar$, the correction to $E_{111} \simeq \epsilon_{111} = 3\hbar\omega_{2}$ due to interaction with state $|n_1^{\prime}n_2^{\prime}n_3^{\prime}\rangle$ is $\Delta E_{111}^{(n_1^{\prime}n_2^{\prime}n_3^{\prime})} \equiv [\epsilon_{n_1^{\prime}n_2^{\prime}n_3^{\prime}}-\epsilon_{111} \pm \sqrt{(\epsilon_{n_1^{\prime}n_2^{\prime}n_3^{\prime}}-\epsilon_{111})^2 + S_{|111\rangle,|n_1^{\prime}n_2^{\prime}n_3^{\prime}\rangle}^2}]/2$, where $S_{|111\rangle,|n_1^{\prime}n_2^{\prime}n_3^{\prime}\rangle} = 2\langle 111|H|\psi_{n_1^{\prime}n_2^{\prime}n_3^{\prime}}^{dr}\rangle$.

Generalizing the way we constructed other dressed states, the dressed state $|\psi_{n_1^{\prime}n_2^{\prime}n_3^{\prime}}^{dr}\rangle$, where  $|n_1^{\prime}n_2^{\prime}n_3^{\prime}\rangle \in \mathbb{H}_{[210]}$ with $\mathbb{H}_{[210]} \equiv \left\{ |210\rangle,\,|021\rangle,\,|102\rangle,\,|201\rangle,\,|120\rangle,\,|012\rangle,\,|111\rangle \right\}$, is constructed as satisfying the Schr\"odinger equation $\langle n_1n_2n_3|H|\psi_{n_1^{\prime}n_2^{\prime}n_3^{\prime}}^{dr}\rangle = E\langle n_1n_2n_3|\psi_{n_1^{\prime}n_2^{\prime}n_3^{\prime}}^{dr}\rangle$ for all basis elements $|n_1n_2n_3\rangle \notin \mathbb{H}_{[210]} - \left\{|n_1^{\prime}n_2^{\prime}n_3^{\prime}\rangle \right\}$      ; assuming $E \approx \epsilon_{n_1^{\prime}n_2^{\prime}n_3^{\prime}}$. 

To lowest order in perturbation theory within the harmonic approximation, the dressed state $|\psi_{201}^{dr}\rangle$ is:
\begin{eqnarray}
	|\psi_{201}^{dr}\rangle &=& |201\rangle - \frac{\sqrt{3}K_{12}|311\rangle}{\epsilon_{201} - \epsilon_{311}} - K_{13} \Bigg( \frac{\sqrt{2}|100\rangle}{\epsilon_{201} - \epsilon_{100}} {} \nonumber \\ 
	& & - \frac{\sqrt{3}|300\rangle}{\epsilon_{201} - \epsilon_{300}} +  \frac{\sqrt{6}|302\rangle}{\epsilon_{201} - \epsilon_{302}} \Bigg) {} \nonumber \\
	& & - \frac{\sqrt{2}K_{23}|212\rangle}{\epsilon_{201} - \epsilon_{212}}.
\end{eqnarray}
The other dressed states can be obtained by applying two-qubit label permutations onto $|\psi_{201}^{dr}\rangle$. Thus, $S_{|111\rangle,|201\rangle}  = S_{|111\rangle,|021\rangle} \simeq 2\sqrt{2}K_{12}$,  $S_{|111\rangle,|210\rangle} = S_{|111\rangle,|012\rangle} \simeq 2\sqrt{2}K_{13}$, $S_{|111\rangle,|120\rangle} = S_{|111\rangle,|102\rangle} \simeq 2\sqrt{2}K_{23}$.

With $\omega_{1,3}=\omega_2 \mp \Delta\omega$ and the qubit-$i$ relative anharmonicity given by $\alpha_{i} = [\omega_{12}^{(i)}-\omega_i]/\omega_i$, where $\omega_{12}^{(i)}$ is the $|1\rangle-|2\rangle$ transition frequency, we can calculate $E_{111} \simeq \epsilon_{111} + \Delta E_{111}^{(201)} + \Delta E_{111}^{(021)} + \Delta E_{111}^{(210)} + \Delta E_{111}^{(012)} + \Delta E_{111}^{(120)} + \Delta E_{111}^{(102)}$, and then the all-to-all $ZZ$ coupling $\zeta_{111}$ given by Eq.(\ref{eq:zz111}) to obtain Eq.(\ref{eq:zeta111}).

\bibliography{graph_3qubits_references}

\end{document}